\documentclass[prb,aps,twocolumn,floats,superscriptaddress,longbibliography]{revtex4-2}
\usepackage{soul}
\usepackage{manfnt}
\usepackage{xcolor}
\usepackage{multirow}
\usepackage{amsmath}
\usepackage{graphicx}
\graphicspath{{../Figs/}}
\usepackage{amssymb}
\usepackage{epsfig}
\usepackage[utf8]{inputenc}
\usepackage[T1]{fontenc}
\usepackage{newtxtext,newtxmath}
\usepackage{pdfrender}
\newcommand{\gtappr}{{{\lower4pt\hbox{$>$} } \atop \widetilde{ \ \ \ }}}
\newcommand{\ltappr}{{{\lower4pt\hbox{$<$} } \atop \widetilde{ \ \ \ }}}

\usepackage{simplewick}
\usepackage[colorlinks = true]{hyperref}
\def\bk{\textbf{k}}

\def\bea{\begin{eqnarray}}
\def\eea{\end{eqnarray}}

\def\dg{^{\dagger}}
\newcommand{\mr}[1]{\mathrm{#1}}

\newlength{\figwidth}
\newlength{\figwidthwide}
\newlength{\shift}
\newcommand{\fgs}[3]
{\begin{figure*}[tbh]\vspace*{-0cm}\centerline{\includegraphics[width=\figwidthwide]{#1}}\vskip
-0.2cm \caption{#3}\label{#2}\end{figure*}}

\newcommand{\ag}[1]{\textcolor{black}{#1}}
\newcommand{\lhl}[1]{\textcolor{black}{#1}}
\begin{document}

\preprint{APS/123-QED}

\title{Oscillate and Renormalize: Fast Phonons Reshape the Kondo Effect in Flat Band Systems}

\author{Liam L.H. Lau}
\email{liam.lh.lau@physics.rutgers.edu}
\affiliation{Center for Materials Theory, Department of Physics and Astronomy,
Rutgers University, 136 Frelinghuysen Rd., Piscataway, NJ 08854-8019, USA}%
\author{Andreas Gleis}%
\affiliation{Center for Materials Theory, Department of Physics and Astronomy,
Rutgers University, 136 Frelinghuysen Rd., Piscataway, NJ 08854-8019, USA}
\author{Daniel Kaplan}%
\affiliation{Center for Materials Theory, Department of Physics and Astronomy,
Rutgers University, 136 Frelinghuysen Rd., Piscataway, NJ 08854-8019, USA}
\author{Premala Chandra}%
\affiliation{Center for Materials Theory, Department of Physics and Astronomy,
Rutgers University, 136 Frelinghuysen Rd., Piscataway, NJ 08854-8019, USA}
\author{Piers Coleman}%
\affiliation{Center for Materials Theory, Department of Physics and Astronomy,
Rutgers University, 136 Frelinghuysen Rd., Piscataway, NJ 08854-8019, USA}

\date{\today}

\begin{abstract}
We examine the interplay between electron correlations and phonons in an Anderson-Holstein impurity model with an Einstein phonon. When the phonons are slow compared to charge fluctuations (frequency $\omega_0 \ll U/2$, the onsite Coulomb scale $U/2$), we demonstrate analytically that the expected phonon-mediated reduction of interactions is completely suppressed, even in the strong coupling regime. This suppression arises from the oscillator's inability to respond to rapid charge fluctuations, manifested as a compensation effect between the polaronic cloud and the excited-state phonons associated with valence fluctuations. We identify a novel frozen mixed valence phase, above a threshold dimensionless electron-phonon coupling $\alpha^*$ when the phonons are slow, where the static phonon cloud locks the impurity into specific valence configurations, potentially explaining the puzzling coexistence of mixed valence behavior and insulating properties in materials like rust. Conversely, when the phonon is fast ($\omega_0 \gtrsim U/2$), the system exhibits conventional polaronic behavior with renormalized onsite interactions effectively $U_{\text{eff}}$ due to phonon mediated attraction, with additional satellite features in the local spectral function due to phonon excitations. Using numerical renormalization group (NRG) calculations, a fully dynamic renormalization technique, we confirm these 
behaviors in both regimes. These findings have important implications for strongly correlated systems where phonon energy scales may be comparable to the Coulomb scale, such as in twisted bilayer graphene, necessitating careful consideration of interaction renormalizations in theoretical models.
\end{abstract}

\maketitle

\section{Introduction} 

Electron-phonon coupling is an important source of fluctuations, leading to structural, electrical, and magnetic instabilities in crystals. Chiefly, it is responsible for superconducting and charge-density wave phase transitions in Fermi liquids. Traditionally, phonons have been considered too slow to induce any Coulomb scale renormalization in strongly correlated systems, as the bare phonon frequency $\omega_0$ is considerably lower than the characteristic charge fluctuation scales of the electronic fluid, notably the interaction scale $U$ and the hybridization $V$. Physically, this means that slow phonons do not have time to respond to the fast charge fluctuations. In the context of the Anderson model, this was studied by \cite{sherrington_ionic_1976, riseborough_strong_1987, hewson1979, hewsonnewns1980, hewson_scaling_1981, hewson_numerical_2001, hewson_numerical_2010,monreal_equation_2009, laakso_functional_2014, eidelstein} showing that for values of $U, V$ relevant for correlated systems such as traditional mixed valent and heavy fermion materials \cite{revhf1, revhf2, revhf3, revhf4, revhf5, Coleman2007}, including $\textrm{Eu}_3\textrm{S}_4$ \cite{berkooz1968}, CeRu$_2$ and  CeIr$_2$~\cite{allen83,gunnarson83}, the phonon contribution is largely perturbative \cite{hewson1979}. However, in $\textrm{YbBAl}_4$, where electrons become slow degrees of freedom compared to the phonons near a quantum criticality point, a polaronic response emerges \cite{kobayashi23}, raising new questions about the importance of electron-phonon coupling when the phonons become fast compared to charge fluctuations.

Recent interest in strongly correlated electron materials has turned to flat-band systems, including gate-tunable moir\'e graphene and dichalcogenide materials,  in which the electronic band-widths
are considerably smaller than the characteristic phonon frequencies \cite{Mak23, guerci23, chen_strong_2024, birkbeck_measuring_2024, cliukphonons, zhu2024, kwan2024, zhida1, zhida2, xidaiphonons}.  In these materials, the phonons are fast degrees of freedom, able to relax in response to the charge fluctuations of the narrow band.   
In particular, experiments \cite{birkbeck_measuring_2024, chen_strong_2024} on twisted bilayer graphene (TBG) show that the flat band in TBG is strongly coupled to an optical $K$ phonon at an energy of $\omega \sim 150 \textrm{meV}$, with strong and pronounced signatures in replica band spectra; these are probed via $\mu$-ARPES imaging of the band structure of the phonon-electron interacting system. A lingering question is therefore how phonons modify the interactions between the flat-band electrons. The effective interaction scale $U_\textrm{eff}$ will then have significant implications for the existence of strong correlations in this system, with experiments showing that aligning the sample to hexagonal Boron Nitride (hBN) causes replica bands to disappear, hinting at the important interplay between flat bands, phonons, and symmetries in the material. 
\figwidthwide=\textwidth
\fgs{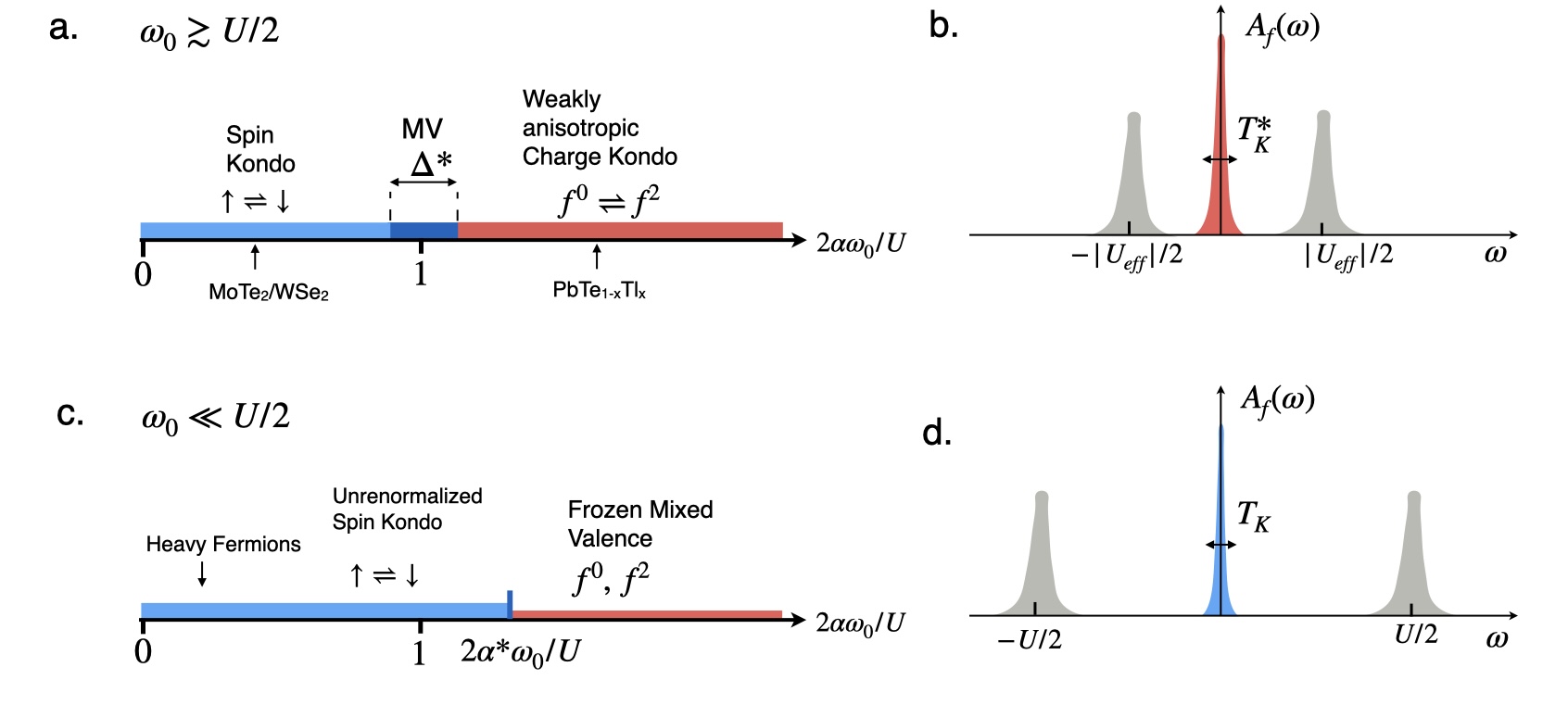}{fig1}{Phase diagram for the Anderson-Holstein model as the reduction in the effective interaction  $U-U_{eff}= 2\alpha \omega_0$.   Two limits are considered: a) Fast phonon ($\omega_0 \gg U/2$) in which increasing electron-phonon coupling $\alpha$ results in a cross-over from spin to charge Kondo effect.  Arrows denote schematically the location of the  Kondo lattice moir\'e chalcogenides MoSe$_2$/WSe$_{2}$ \cite{Mak23, guerci23} (Kondo effect with reduced $U_{\rm eff}$) and the charge Kondo system PbTe$_{1-x}$Tl$_x$\cite{fisher94}. b) shows f-electron spectral function with upper and lower side bands at $\pm |U_{\rm eff}|/2$ and renormalized Kondo temperature $T_K^*$ as the width of the central Kondo resonance for both spin and charge Kondo regimes for both spin and charge Kondo regimes,   c) Slow phonon ($\omega_0\ll U/2$), showing d) f-electron spectral function for the spin Kondo regime, with upper and lower side bands at unrenormalized $\pm U/2$ and Kondo temperature $T_K$ as the width of the central Kondo resonance.}

Motivated by these considerations, here we revisit 
the effect of phonons on electronic correlations using an Anderson-Holstein impurity Hamiltonian $H=H_{A}+H_H$, involving a mixed valent atom  coupled to a single optical phonon, where
\begin{eqnarray}\label{eq:andersonHammy}
    H_{\text{A}} &=&\sum_{\bk \sigma} \epsilon_{\bk} c\dg_{\bk \sigma} c_{\bk \sigma} + V\sum_{ \sigma} \left( c\dg_{\sigma} f_{\sigma} +  {\rm H.c.}\right) + \frac{U}{2} (n_f -1)^2\cr &&
\end{eqnarray}
is the symmetric Anderson model that describes a localized electron, created by $f\dg_{\sigma}$ with spin component $\sigma= \pm 1$
hybridized with a conduction sea;  $U$ is the onsite Coulomb interaction;   $c\dg_{\sigma}$ creates a conduction electron at the origin with spin component $\sigma$.  
\begin{equation}\label{eq:andersonholsteinHammy}
    H_H = \omega_0 b\dg b + g\left(b + b\dg\right) (n_f - 1)
\end{equation}
describes an Einstein phonon of frequency $\omega_0$ with phonon creation operator $b^{\dagger}$, coupled linearly to the deviations in local impurity charge from half filling $(n_f - 1)$ via the coupling constant $g$.  In terms of these quantities, we can define a dimensionless electron-phonon coupling strength $\alpha = g^2/\omega_0^2$, so that $g= \omega_0 \sqrt{\alpha}$. Physically, the impurity charge couples to the phonon position operator $x = (b + b\dg)/\sqrt{2  \omega_0}$. This interaction displaces the ionic equilibrium positions in opposite directions depending on whether the impurity is empty or doubly occupied.

We summarize the effect of the electron-phonon coupling on the zero temperature phase diagram and the local spectral function of the impurity electrons in Fig. \ref{fig1}. We begin by discussing the energy scales and the current understanding.
The case $\alpha=0$ describes the conventional repulsive $(U > 0)$ Anderson model. The non-interacting Anderson model describes a simple resonance level of width $\Delta$ given by Fermi's golden rule, $\Delta = \pi \rho V^2 $, where $\rho$ is the density of conduction electron states.  When $U$ becomes substantially larger than $\Delta$, charge fluctuations are suppressed, and the spectral function of the f-electron divides into a lower and upper Hubbard peak at energies $\pm U/2$, plus a central "Kondo resonance" of width (as shown in Fig. \ref{fig1}d) given by the Kondo temperature
\bea\label{eq:KT}
T_K = \sqrt{U \Delta}e^{- \frac{\pi U}{8\Delta}}
\eea
At temperatures larger than the Kondo effect, the model describes a decoupled local moment, but at temperatures $T < T_K$, the local moment is screened by the conduction electrons, giving rise to the development of Fermi liquid. \lhl{Above the Kondo temperature $T_K$, scattering is inelastic and lacks a well-defined real phase shift $\delta$, while below $T_K$, it becomes elastic with a phase shift of $\pi/2$ signaling the formation of an s-wave Kondo singlet bound state.}

The principle effect of the phonons is to induce an attraction between the f-electrons, reducing the onsite Coulomb interaction according to the Fr\" ohlich interaction\cite{AndersonnegativeU}
\begin{equation}\label{eq:renormint}
U\rightarrow U_{\text{eff}} = U -\frac{2 g^2}{\omega_0}=U - 2 \alpha \omega_0.
\end{equation}
Additionally, if the frequency of the phonon dynamics is faster than that of the valence fluctuations, i.e.  $\omega_0 \gg \Delta$, then the ionic equilibrium positions relax every time an electron hops in or out of the impurity which causes the bare hybridization width to exponentially renormalize \cite{sherrington_ionic_1976, riseborough_strong_1987},
\begin{equation}\label{eq:renormdelta}
    \Delta \rightarrow\Delta_{\text{eff}} = \Delta \exp{\left[-\alpha \coth{(\beta\omega_0/2)}\right]} \xrightarrow[]{T = 0} \Delta e^{-\alpha}.
\end{equation}
The renormalizations \eqref{eq:renormint}  and \eqref{eq:renormdelta} lead to a corresponding renormalization of the  Kondo temperature \eqref{eq:KT}
\begin{equation}\label{eq:renormkt}
    T^*_K = \sqrt{|U_{\text{eff}}| \Delta_{\text{eff}}}\; e^{- \frac{\pi |U_{\text{eff}}|}{8\Delta_{\text{eff}}}},
\end{equation}
resulting in a corresponding reduction in the Kondo resonance width in the f-electron spectral function(Fig. \ref{fig1}b). 

\begin{figure}
    \centering
    \includegraphics[width=1.0\linewidth]{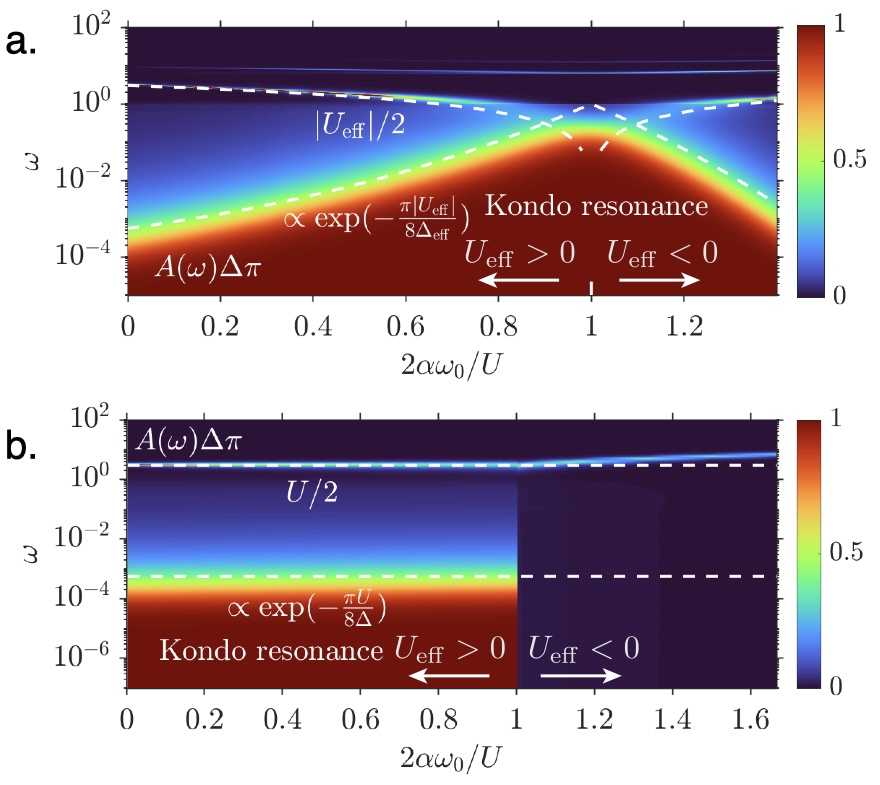}
    \caption{Numerical renormalization group (NRG) results showing the $\alpha$-dependence of the impurity spectral function for the: (a) fast phonon scenario ($U = \omega_0 = 6$ and bare hybridization width $\Delta/\pi = 0.1$ with renormalized $\Delta_{\mr{eff}} = e^{-\alpha} \Delta$, $U_{\mr{eff}} = U - 2\alpha\omega_0$) displaying renormalized excitation energy $U_{\text{eff}}/2 = U/2 - \alpha \omega_0$ and renormalized Kondo resonance; and (b) slow phonon scenario ($U =6$ and $\Delta/\pi = \omega_0 = 0.1$) where the excitation energy remains at $U/2$ with an unrenormalized Kondo resonance until the dimensionless electron-phonon coupling exceeds threshold $\alpha^*$, where $2\alpha^*\omega_0/U > 1$, causing the the Kondo resonance to disappear.}
    \label{fig:nrgspectral}
\end{figure}

As the strength of the electron-phonon coupling $\alpha$ rises the interaction \eqref{eq:renormint} between the f-electrons ultimately changes sign, causing a transition from a positive $U$ Anderson model, described by spin Kondo effect, via a  weakly interacting  mixed valent Fermi liquid, where the renormalized hybridization width $\Delta_{\mr{eff}}$ is larger than the effective onsite interaction $U_{\text{eff}}$, into a region governed by a negative $U$\cite{AndersonnegativeU}.The atomic limit ($V=0$) is described by a ground-state spin doublet  $(\vert f^1:\uparrow\rangle, \vert f^1:\downarrow\rangle)$ for positive $U_{\text{eff}}>0$, and a charge doublet of singly and doubly occupied f-states, $(\vert f^0\rangle, \vert  f^2\rangle)$  for $U_{\text{eff}}<0$\cite{AndersonnegativeU}. For small hybridization $V>0$, the resulting fluctuations within these doublets give rise to spin or charge Kondo effects, as follows
\begin{itemize}
\item $U_{\text{eff}}>0$: Spin Kondo effect $\vert \uparrow \rangle \rightleftharpoons \vert \downarrow\rangle$.
\item $U_{\text{eff}}<0$: Charge Kondo effect\cite{AndersonnegativeU,Taraphder91}, $\vert f^0 \rangle \rightleftharpoons \vert f^2\rangle$.
\end{itemize}
Whereas the former results in a magnetically polarizable Fermi liquid, the second leads to an electrically polarizable Fermi liquid with a large charge susceptibility. The negative $U$ region is associated with soft pair fluctuations and a corresponding charge Kondo effect \cite{Taraphder91,fisher94} (see schematic Fig. \ref{fig1}a and a summary of detailed numerical renormalization group calculations shown in Fig. \ref{fig:nrgspectral}a). 
 The concept of negative $U$ centers plays an important role in the understanding of vacancies in amorphous silicon \cite{AndersonnegativeU,Schluter}.
Negative $U$ physics has also been observed in thalium-doped lead telluride PbTe$_{1-x}$Tl$_{x}$\cite{fisher94}, giving rise to a competition between a charge Kondo effect and superconductivity\cite{dzeroSC}. 

An analysis of the corresponding Fermi liquid model \cite{yamadalocal,yosidalocal} shows that the linear specific heat of the impurity $C= \gamma T$ contains a charge and a spin component. Normalizing the specific heat, spin, and charge susceptibilities with respect to their respective non-interacting  ($U=0$) values, $\tilde \gamma = (\gamma/\gamma_{0})$, $\tilde\chi_s = (\chi_s/\chi^{0}_{s})$ and $\tilde\chi_c = (\chi_c/\chi^{0}_{c})$ , the Yamada-Yosida relation\cite{yamadalocal,yosidalocal} states that 
\bea
2 \tilde\gamma = \tilde\chi_s + \tilde\chi_c
\eea
reflecting the fact that the linear specific heat contains a spin and charge component. Rewriting this expression in terms of the spin Wilson ratio $W_s= \tilde \chi_s/\tilde \gamma $ and charge Wilson ratio $W_c =\tilde \chi_c/\tilde \gamma$ gives $W_s+W_c=2$. In the 
spin Kondo effect, the charge susceptibility of the Fermi liquid vanishes, so $W_s=2$; by contrast, in the charge Kondo effect, where the spin susceptibility vanishes, $W_c=2$. The nature of the transition between the spin and charge Kondo limits is of particular interest.

This paper addresses these issues with a combination of perturbative and state-of-the-art numerical renormalization group. 
One of the key points to emerge from our work is that it is the phonon time scales and not the strength of the electron-phonon interaction that determines whether phonons influence the electrons. In our discussion, we assume that $U$ is large compared to $\Delta$, so prior to coupling
to the phonons, the electrons are in the Spin Kondo regime. 

Two limiting cases of the Anderson-Holstein model illustrate this conclusion: fast phonons, where $\omega_0 \gtrsim U/2$, and slow phonons where  $\omega_0 \ll U/2$. In the "fast" phonon limit, the system evolves from a spin Kondo phase to a weakly anisotropic charge Kondo regime as a function of increasing electron-phonon coupling $\alpha$, passing through the point $\alpha =\alpha_0$,  where the effective onsite interaction vanishes $U_{\text{eff}} = U - 2\alpha_0 \omega_0 = 0$ and the charge and spin susceptibilities,  $\tilde \chi_s = \tilde \chi_c= \tilde \gamma$ are equal, as in a non-interacting Anderson model (see the schematic in Fig. \ref{fig1}a and corresponding numerical renormalization group results Fig. \ref{fig:nrgspectral}a which shows the renormalized excitation scale and Kondo scale in both the spin Kondo and charge Kondo regimes). 
\begin{figure}[t!]
    \centering
    \includegraphics[width=1.0\linewidth]{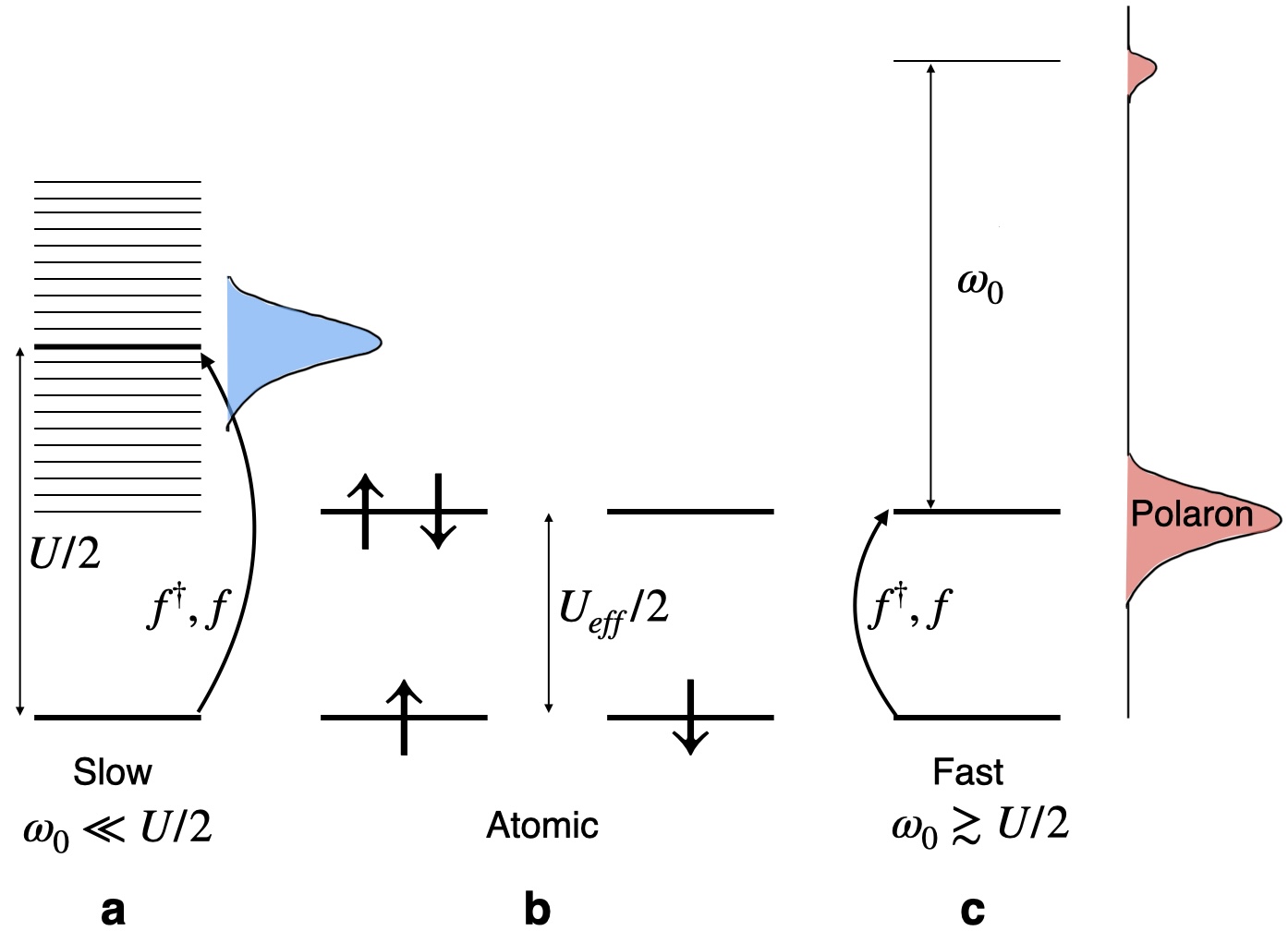}
    \caption{Schematic showing the excitation spectrum of the: a) Slow phonon ($\omega_0 \ll U/2$) with a tower of phonon excitations spaced by $\omega_0$, a subtle compensation effect between the most probable phonon excitation energy $\alpha \omega_0$ keeps the excitation energy to add or remove an electron to the impurity at the unrenormalized value $U/2$ because the phonons are too slow to screen the onsite interaction; b) Atomic limit of the Anderson-Holstein model, where the formation of the polaron is implicitly assumed; and the c) Fast phonon ($\omega_0 \gtrsim U/2$) with additional phonon satellites spaced every $\omega_0$. The dominant excitation energy when adding or removing an electron from the impurity is $U_{\text{eff}}/2$, which includes a polaronic reduction because the phonon cloud can react. }
    \label{fig:bottomlineschematic}
\end{figure}

By contrast, "slow" phonons ($\omega_0 \ll U/2$) cannot respond to high frequency valence fluctuations. A rather subtle compensation takes place in the physics so that the static shift in $U$ seen in the atomic limit is entirely compensated by the slow phonon relaxation,  so that the electrons are blissfully unaware of the phonons, experiencing an unrenormalized $U$. The low energy physics is therefore that of a spin Kondo model. However, above a threshold dimensionless electron-phonon coupling strength $\alpha^*$, the ground state energy of the static charge configurations drops below the ground-state energy of the positive $U$ Anderson model, and a transition occurs from a spin Kondo effect into a frozen mixed valence state  $(f^0, f^2)$ with no Kondo effect (see schematic in Fig. \ref{fig1}c and corresponding numerical renormalization group results in
Fig. \ref{fig:nrgspectral}b which confirm the disappearance of the Kondo resonance above the threshold   coupling $\alpha^*$). In the slow limit, where $\omega_0$ is small but $\alpha$ is large, tunneling between the valence configurations $f^0$ and $f^2$ is exponentially suppressed by a factor of $e^{-2\alpha}$, causing the charge Kondo effect to freeze out into static valence configurations reminiscent of iron in rust \cite{rustsmallpolaron}. 

\lhl{In approaching these questions, we draw inspiration from Hewson and Meyer's foundational work \cite{hewson_numerical_2001}, which combined analytical and numerical renormalization group (NRG) methods to gain valuable insight into the interplay between electron-phonon coupling and electronic correlations. While they focused on systems with magnetic impurities, their simulations used parameters ($\omega_0 = U/2 \sim \pi \Delta$) that place their work in what we identify as the fast phonon regime, where the high energy spectral features associated with atomic correlations are renormalized as the phonon coupling is increased. This was understood using perturbation theory in their work. Our study is an extension of their work by developing a more comprehensive phase diagram, particularly revisiting the slow phonon case ($\omega_0 \ll U/2$) and demonstrating that high-energy atomic features remain unaffected, unlike in the fast phonon case described by Hewson and Meyer. We present a unified framework encompassing fast and slow phonon regimes across weak and strong electron-phonon coupling limits. For the slow phonon case, we explore the previously underappreciated negative $U_{\text{eff}} = U - 2\alpha \omega_0$ region and discuss the potential relevance of the fast phonon case for two dimensional van der Waals materials.}

\lhl{The distinction between adiabatic (slow phonons) and antiadiabatic (fast phonons) regimes in the Anderson-Holstein problem has been well-established in the literature. This dichotomy appears in various contexts, including mixed-valent heavy fermion compounds \cite{sherrington_ionic_1976, riseborough_strong_1987} and phonon-assisted tunneling \cite{eidelstein}, which compared the phonon frequency $\omega_0$ with a measure of the electronic movement- the bare hybridization width $\Delta$ and the bare tunneling rate, respectively. Notably, Hewson and Newns \cite{hewson1979,hewsonnewns1980} investigated the polaronic reduction of the virtual bound state width, demonstrating that this effect emerges when the relaxation energy $\alpha \omega_0$ exceeds the conduction bandwidth, particularly when the phonon frequency $\omega_0$ is larger than the bandwidth. While they noted that typical heavy fermion materials don't reach this parameter range (thus electronic hopping into and out of a given f-level isn't polaronically reduced), they found that inter-site valency migration experiences polaronic reduction. Our work extends these findings by showing that the effective Anderson model undergoes renormalization only when the phonon frequency exceeds the Coulomb scale ($\omega_0 > U/2$), which consequently renormalizes both the Kondo temperature \eqref{eq:renormkt} and the virtual bound state width.}

This paper is structured as follows. Section \ref{sec:model}\ introduces the key energy scales of the Anderson-Holstein impurity model from the atomic limit, rewriting the Hamiltonian in a translated basis where the phonon oscillator's equilibrium position has shifted due to impurity coupling. In \ref{sec:analytic}, we develop an analytic approach to the static renormalization of these scales when phonons are either slow or fast compared to charge fluctuation timescales- noting that the onsite interaction is unrenormalized with slow phonons because the phonons cannot react in time to do anything, and renormalized with fast phonons due to the formation of onsite polarons. In  \ref{sec:NRGsec}, we present numerical renormalization group (NRG) calculations for dynamic properties (spectral functions) and thermodynamic properties (impurity entropy contribution, spin Wilson ratios, and spin/charge susceptibilities), which confirm our analytical physical picture. The data that support the findings of this article are openly available \cite{lau2025dataset}. We conclude with a summary and outlook in section \ref{sec:discussion}.

\section{Model}\label{sec:model}

We consider the symmetric Anderson-Holstein impurity model, as discussed in the introduction,
\begin{equation}\label{eq:andersonholsteinHammy}
    H = H_A + H_H
\end{equation}
with an electron-phonon coupling $g = \sqrt{\alpha} \omega_0$. For the symmetric single impurity Anderson model $H_A$, the non-interacting scenario describes a resonance level with an inverse lifetime given by Fermi's golden rule, $\Delta = \pi \rho V^2$, otherwise known in the literature as the resonance width. For the atomic limit ($V = 0$) of the symmetric single impurity Anderson model $H_A$ with repulsive onsite interaction $U$, the ground state of the impurity is a singly occupied state $\vert \sigma \rangle$ with spin $\sigma$, with an energetic cost to remove or add an electron,
\begin{equation}
    \Delta E^A_\pm = \frac{U}{2} > 0.
\end{equation}
Since the ground state manifold is a doubly degenerate and spinful, it is known as the local moment sector in the literature. 

When the local electron-phonon coupling to an optical Einstein model with constant frequency $\omega_0$ is turned on, it is still useful to consider the atomic limit,
\begin{equation}\label{eq:atomicAH}
    H_{\text{atomic}} = \omega_0 b\dg b + \sqrt{\alpha} \omega_0 \left(b + b\dg\right)(n_f - 1) + \frac{U}{2} (n_f-1)^2.
\end{equation}
Completing the square for the phonons, 
\begin{eqnarray}\label{eq:completesquare}
    H_{\text{atomic}} &=& \omega_0 (b\dg+ \sqrt{\alpha} (n_f-1))( b + \sqrt{\alpha} (n_f-1)) \cr &+& \frac{\left(U - 2\alpha \omega_0 \right)}{2} (n_f-1)^2,
\end{eqnarray}
can be understood as a displaced harmonic oscillator coupled to the impurity,
\begin{equation}
    H_{\text{atomic}} = \frac{\hat{p}^2}{2} + \frac{ \omega_0^2}{2}(\hat{x}-x_0)^2 + \frac{(U - 2\alpha \omega_0)}{2}(n_f - 1)^2,
\end{equation}
where the position and momentum operators of the phonon oscillator are $\hat{x} = (b + b\dg)/\sqrt{2  \omega_0}$ and $\hat{p} = i\sqrt{\omega_0/2}(b\dg - b) $ respectively. The location of the displaced  oscillator,
\begin{equation}\label{eq:displacedeqm}
   x_0[n_f] = \sqrt{\frac{2\alpha}{\omega_0}} (1-n_f),
\end{equation}
depends on whether the impurity is empty or doubly occupied. The phonon-induced attraction $-\alpha \omega_0(n_f-1)^2$ can be interpreted as just a shift in the reference energy $-\omega_0^2 x_0^2/2$, which assumes the phonon oscillator is displaced by $x_0$.

The position operator $x$ of the phonon oscillator can be redefined to describe the fluctuations around the equilibrium position $x\rightarrow \tilde{x}+x_0$. Formally, this is accomplished using the unitary transformation $\hat x \rightarrow e^{i \hat p x_0}\hat x e^{-i \hat p x_0}$, where 
\begin{equation}
   \exp{(-i \hat{p} x_0)} = \exp{\left( \sqrt{\alpha}(b - b\dg)(n_f - 1)\right)}.
\end{equation}
is the translation operator. 
The action of this unitary translation operator on the atomic Anderson-Holstein model \eqref{eq:atomicAH} is to shift the harmonic oscillator (known in the literature as the Lang-Firsov transformation- Appendix \ref{sec:details}), 
\begin{equation}
    e^{i \hat{p} x_0} H_{\text{atomic}} e^{-i\hat{p}x_0} =  \omega_0 \tilde{b} \dg \tilde{b} + \frac{U_{\text{eff}}}{2}(n_f-1)^2,
\end{equation}
where the effective onsite interaction is $U_{\text{eff}} = U - 2\alpha \omega_0$, the shifted phonon operator is,
\begin{equation}\label{eq:shiftedphonon}
   \tilde{b}\dg = b\dg + \sqrt{\alpha} (n_f -1),
\end{equation}
and the linear electron-phonon coupling term is eliminated. Here, $n_f$ is the number operator for the polaron.  The original f-creation operator now transforms to 
\begin{eqnarray}\label{eq:polaron}
   f\dg_{\sigma} \rightarrow e^{i\hat{p} x_0} f\dg_\sigma e^{-i \hat{p}x_0} =  D {\tilde f}\dg_{\sigma},
\end{eqnarray}
where $D = \exp{\left( \sqrt{\alpha} (b\dg-b )\right)} =\exp{\left( \sqrt{\alpha} (\tilde{b}\dg-\tilde{b} )\right)} $.  Here,  ${\tilde f}\dg_{\sigma}$ creates the polaron, while  $D$ undresses the impurity from the polaron cloud to ensure $f\dg_\sigma$ creates the bare impurity electron.  Under this unitary transformation, the charge is unaltered, i.e $n_{\tilde f}=n_{f}$.  Henceforth, we will drop the tilde on the f operators, with the understanding that in the transformed Hamiltonian, the $f\dg $ are polaron creation operators. 

The displaced phonon ground state is annihilated by the shifted phonon operator, $\tilde{b} \vert \tilde 0 \rangle = 0$. For a singly occupied impurity ($n_f = 1$), the phonon ground state is the untranslated phonon vacuum $\vert 0 \rangle$. However, the phonon ground states for the empty and doubly occupied impurity involve opposite displacements from equilibrium $x_0$ \eqref{eq:displacedeqm}. In the original basis, the empty and doubly occupied ground-states are given by
\bea
\vert {\tilde 0}^+ \rangle &=& e^{-i \hat{p} x_0[n_f=0]} \vert {0} \rangle = e^{\sqrt{\alpha}(b\dg- b)} \vert {0}\rangle, \quad (n_f=0)\cr
\vert {\tilde 0}^- \rangle &=& e^{-i \hat{p} x_0[n_f=2]} \vert {0} \rangle = e^{-\sqrt{\alpha}(b\dg- b)} \vert {0}\rangle, \ \ (n_f=2).
\eea

Fig. \ref{fig:bottomlineschematic}b summarizes the key impurity energy scales from the atomic model when the displaced phonon $\tilde{b}$ is in the ground state $\vert \tilde{0} \rangle$. Onsite interaction $U$ is renormalized to an effective value  $U_{\text{eff}} = U - 2 \alpha \omega_0$ due to polaron formation. When an electron is added to or removed from the singly occupied impurity without a phonon cloud, it creates  $\tilde{n}_b = \tilde{b}\dg \tilde{b}$ phonon excitations, the energetic cost to do so is,
\begin{equation}\label{eq:AHatomicexciation}
   \Delta E^{AH}_\pm = \frac{U_{\text{eff}}}{2} + \tilde{n}_b \omega_0,
\end{equation}
corresponding to the ladder of states in Figures \ref{fig:bottomlineschematic}a and \ref{fig:bottomlineschematic}c spaced by $\omega_0 \ll U/2$ and $\omega_0 \gtrsim U/2$ respectively.

While energy minimization suggests this process should involve no phonon excitations ($\tilde{n}_b = 0$) on top of the polaron state, the actual physics depends on the phonon frequency $\omega_0$ (Fig. \ref{fig:bottomlineschematic}a and Fig. \ref{fig:bottomlineschematic}c), and will be discussed in detail in the following section. The bottom line is, if $\omega_0$ is slow compared to the bare onsite interaction $U/2$, the phonon cloud cannot respond quickly enough to dress the impurity, leaving the single impurity Anderson model unrenormalized despite strong electron-phonon coupling. Conversely, if phonons are fast ($\omega_0 \gtrsim U/2$), the cloud can effectively respond and renormalize the model. 

Turning on the hybridization $V$, the full Hamiltonian \eqref{eq:andersonholsteinHammy} in the translated basis $\tilde{H} = e^{i\hat{p} x_0} H e^{-i\hat{p}x_0}$ is,
\begin{eqnarray}
   \hskip -0.1in\tilde{H} = \omega_0 \tilde{b}\dg \tilde{b} + \frac{U_{\text{eff}}}{2}(n_f - 1)^2 + V \sum_\sigma \left(c\dg_\sigma f_\sigma D\dg + \text{H.c.} \right).
\end{eqnarray}

\begin{figure}[t]
    \centering
    \includegraphics[width=0.9\linewidth]{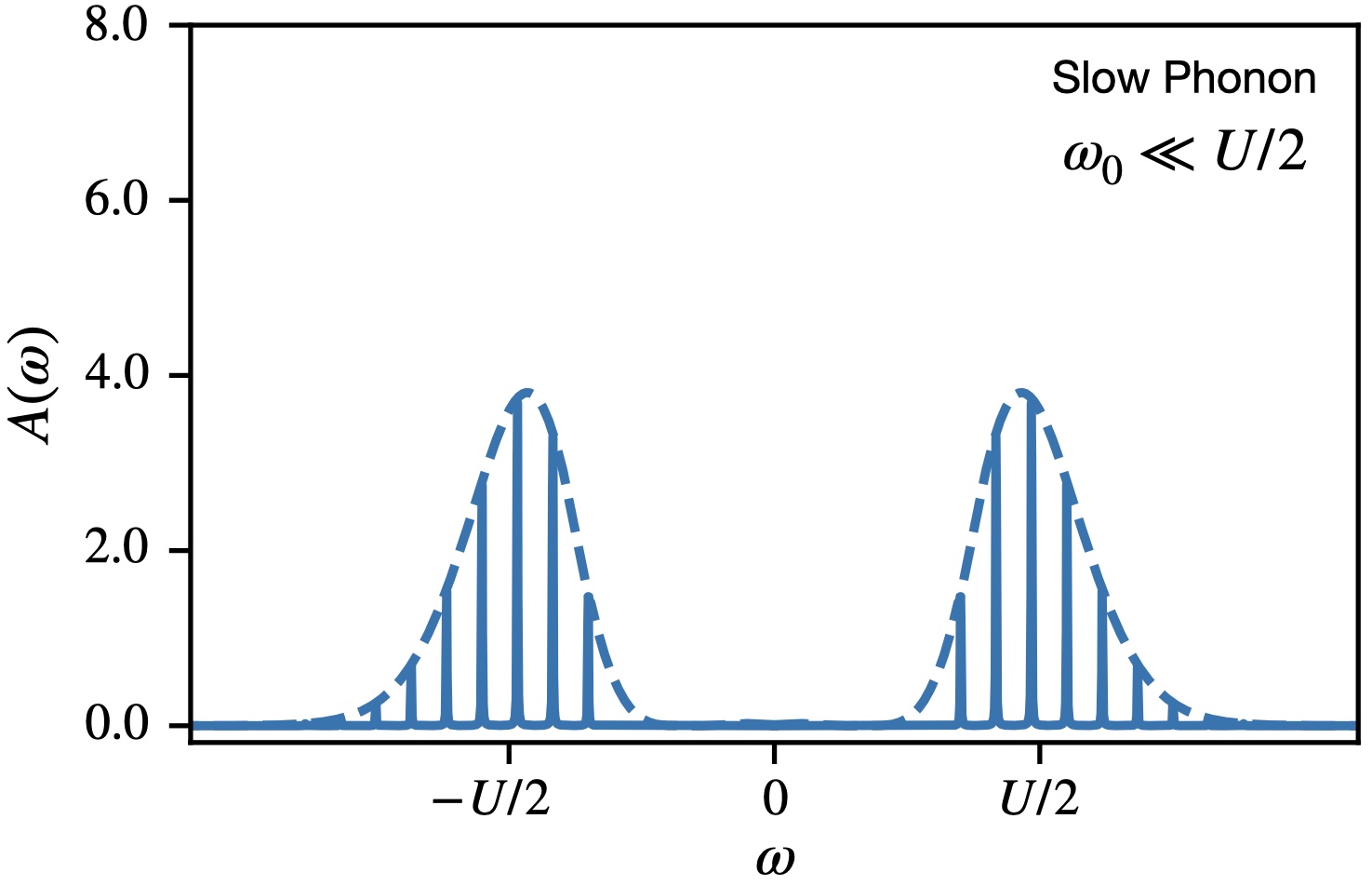}
    \caption{Impurity spectral function peaks and envelope function for the Anderson-Holstein model in the atomic limit, calculated using the Poisson distribution for the number of excited phonons ($n_b$). This figure shows the case where phonons are slow compared to the onsite interaction ($\omega_0 \ll U/2$). The maximum of the distribution occurs at $\pm U/2$, matching the upper/lower Hubbard band positions for the bare single impurity Anderson model. The key energy scales remain unaffected by coupling to a slow phonon mode because the phonon cloud cannot respond in time to the charge fluctuations.}
    \label{fig:slowspectral}
\end{figure}

\section{Analytic Approach to Static Renormalizations}\label{sec:analytic}
To analytically explore this phonon frequency dependence, we examine the atomic Hamiltonian \eqref{eq:completesquare} in both slow ($\omega_0 \ll U/2$) and fast phonon ($\omega_0 \gtrsim U/2$) regimes. 

When an electron is removed or added from the impurity, the resulting phonon clouds are driven out of equilibrium, and are no longer in the ground state. The state created by removing an electron from the singly occupied ground-state is
\begin{eqnarray}
    \vert {-}\rangle = D\dg \vert \tilde{0} \rangle = \exp{\left(- \sqrt{\alpha}(b\dg - b)\right)} \vert \tilde{0}\rangle,
\end{eqnarray}
while the state created by adding an electron to this state is 
\begin{eqnarray}
    \vert {+}\rangle = D \vert \tilde{0} \rangle = \exp{\left(+ \sqrt{\alpha}(b\dg - b)\right)} \vert \tilde{0} \rangle.
\end{eqnarray}

To see that these states contain a mixture of excited state phonons, consider the 
the matrix element between $\vert{\pm} \rangle$ and the state $\vert \tilde{n}_b\rangle $, with $\tilde{n}_b$ excited phonons,
\begin{equation}\label{eq:matrixelement}
\langle \tilde{n}_b\vert {\pm} \rangle = \langle \tilde{0} \vert \frac{\tilde{b}^n}{\sqrt{n!}} e^{\pm \sqrt{\alpha}\tilde{b}\dg}e^{\mp\sqrt{\alpha}\tilde{b}}e^{-\frac{\alpha}{2}}\vert \tilde{0}\rangle = \frac{(\pm\sqrt{\alpha})^n}{\sqrt{n!}}e^{-\frac{\alpha}{2}},
\end{equation}
where we have used the Baker–Campbell–Hausdorff identity $e^{A+B} = e^Ae^B e^{-[A, B]/2}$ to normal-order the coherent state operator $D = e^{\sqrt{\alpha}\tilde{b}\dg}e^{-\sqrt{\alpha}\tilde{b}}e^{-\frac{\alpha}{2}}$. Notice the interesting even-odd alternation in the sign of the overlap 
with the empty state $\vert -\rangle$. 

The probability of creating a state with $\tilde{n}_b$ excited phonons when an electron is added  or removed from the singly occupied impurity state is determined by the squared  matrix element between $\vert \tilde{n}_b\rangle$ and the phonon coherent states $\vert {\pm}\rangle$,
\begin{equation}\label{eq:poisson}
    P_{\alpha}(\tilde{n}_b) = \vert \langle \tilde{n}_b \vert {\pm} \rangle\vert^2 = \frac{ \alpha^{\tilde{n}_b}}{\tilde{n}_b!}e^{-\alpha} ,
\end{equation}
which corresponds to a Poisson distribution centered around a mean-phonon number $\langle \tilde n_b\rangle = \alpha$ with standard deviation $\sqrt{\langle \delta \tilde n_b^2 \rangle}= \sqrt{\alpha}$. In the slow phonon limit where $\alpha \rightarrow \infty $ this distribution function becomes a normalized delta-function centered at $\tilde n_b=\alpha$ excited phonons $P_{\alpha}(\tilde n_b)\sim \delta (\tilde n_b - \alpha)$.  The energy of the doubly occupied or empty states with $\tilde n_b=\alpha $, 
\bea E= \frac{U_{\text{eff}}}{2} + \omega_0 \langle \tilde n_b\rangle  = \frac{U}{2},\eea
contains the unrenormalized Coulomb interaction $U$.   This is in essence, a consequence of Ehrenfest's theorem.  The slow phonon limit represented by large $\alpha$, small $\omega$ is, in essence, a classical limit, in which the polaron clouds behave as an off-resonance classical oscillator that does not respond to valence fluctuations.   We can thus derive a {\sl phonon compensation theorem}:  that the compensation between the polaronic reduction of $U_{\text{eff}} = U - 2 \alpha \omega_0$ and the number of phonons created by the valence fluctuation becomes perfect in the slow phonon limit, reflecting the associated inability of the classical polaron cloud to respond to the fast electrons and renormalize the fast Anderson dynamics.  

We can see this another way by looking at the f-electron spectral function in the atomic limit. Since we know the energy to add or remove an electron from the singly occupied impurity is $\Delta E^{AH}_\pm$ \eqref{eq:AHatomicexciation} and we also know that the probability of exciting $\tilde{n}_b$ phonons during this process \eqref{eq:poisson}, we can immediately write down the  impurity Green's function in the Lehmann representation for the symmetric Anderson-Holstein model for the atomic limit when the impurity ground state is singly occupied,
\begin{align}\label{eq:atomicgreensfunc}
      \notag & G^{(AH)}_\sigma (\omega) = \\ \notag & \sum_{\tilde{n}_b =0}^{\infty} \left[\frac{|\langle f^2; \tilde{n}_b \vert f\dg_\sigma D \vert  \sigma; \tilde{0} \rangle|^2}{\omega -\frac{1}{2}U_{\rm eff} - \tilde{n}_b \omega_0} +\frac{|\langle f^0; \tilde{n}_b\vert f_{\sigma} D\dg \vert \sigma; \tilde{0} \rangle|^2}{\omega +\frac{1}{2}U_{\rm eff} + \tilde{n}_b \omega_0}\right] =  \\ &  \frac{e^{-\alpha}}{2} \sum_{\tilde{n}_b =0}^{\infty} \frac{\alpha^{\tilde{n}_b}}{\tilde{n}_b!} \left[\frac{1}{\omega -\frac{1}{2}U_{\rm eff} - \tilde{n}_b \omega_0} +\frac{1}{\omega +\frac{1}{2}U_{\rm eff} + \tilde{n}_b \omega_0}\right],
\end{align}
where $\vert \sigma; \tilde{0} \rangle$ is the singly occupied impurity state where the translated phonon is unoccupied, $\vert f^0; \tilde{n}_b \rangle$ and $\vert f^2; \tilde{n}_b \rangle$ are the empty and doubly occupied impurity states where there are $\tilde{n}_b$ translated phonons occupied, reproducing the result from Hewson and Meyer \cite{hewson_numerical_2001}. The first term in the square brackets is associated with adding an electron into the singly occupied impurity with $\tilde{n}_b$ phonon excitations. In contrast, the second term in the square brackets is removing an electron from the singly occupied impurity with $\tilde{n}_b$ phonon excitations. 

In the following subsections, we demonstrate that the phonon response to charge fluctuations is captured by the peak of the phonon number Poisson distribution \eqref{eq:poisson}. Counter to simple energy minimization arguments, the most probable state after electron addition or removal may contain $\tilde{n}_b \neq 0$ phonon excitations. When phonons are too slow ($\omega_0 \ll U/2$), they cannot respond quickly enough to modify electron interaction physics, captured by a subtle compensation in the excitation energies \eqref{eq:AHatomicexciation} where the peak in the most probable phonon energy keeps the excitation energy unrenormalized (Fig. \ref{fig:bottomlineschematic}a). Conversely, when phonons are sufficiently fast ($\omega_0 \gtrsim U/2$), they renormalize the single impurity Anderson model and generate phonon side peaks (Fig. \ref{fig:bottomlineschematic}c).

\begin{figure*}[t!]
    \centering
    \includegraphics[width=13cm]{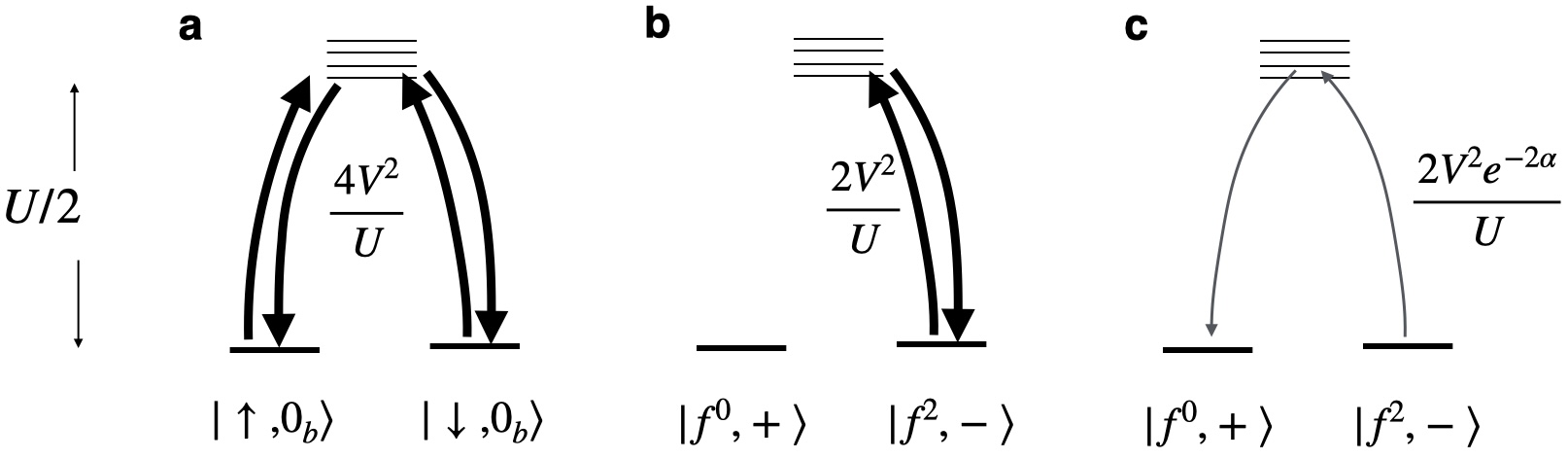}
    \caption{Illustrating the virtual valence fluctuations in the spin Kondo limit a) Kondo spin exchange fluctuations, and charge Kondo limit b) ``Ising''  fluctuations which return to the same valence state are not suppressed by the polaron cloud c) ``Transverse'' fluctuations which skip the valence by two are suppressed by a factor of $e^{-2\alpha} = \langle - \vert + \rangle $ associated with the overlap between the polaron clouds of the $f^0$ and $f^2$ configurations. }
    \label{fig:anisotropyflucs}
\end{figure*}

\subsection{Dynamics in the Slow Phonon Regime}
\label{sec:slowphonons_analytic}

When phonons are slow relative to both the bare hybridization width $\Delta$ and the interaction scale $U/2$, creating an energy hierarchy where $\omega_0 \ll \Delta \ll U/2$, the phonons cannot respond quickly enough to the charge fluctuations. The polaron clouds are classical in this limit, and their frequency is far off-resonance preventing any phonon response. This effectively freezes out phonon dynamics, leaving the single impurity Anderson model unaltered. 

The excitation energy to add or remove an electron from the singly occupied impurity with the phonon cloud is $\Delta E^{AH}_\pm = U_{\text{eff}}/2 + \tilde{n}_b \omega_0$, shown as the tower of states separated by $\omega_0 \ll U/2$ above $U_{\text{eff}}/2$ (Fig. \ref{fig:bottomlineschematic}a), where the occupation of $\tilde{n}_b$ follows the Poisson distribution \eqref{eq:poisson}. For slow phonons where $\omega_0 \ll U/2 $ and the dimensionless electron-phonon coupling is strong $\alpha \gg1$, the Poisson distribution for the probability for phonon excitations Eq. \eqref{eq:poisson} is sharply peaked at $n_b \approx \alpha$ with excitation energy $\alpha \omega_0$, contrary to the naive expectation that no phonons should be excited based on energetic arguments.

Consequently the excitation energy to add or remove an electron from the singly occupied impurity in the slow phonon regime is dominated by the most probable phonon excitation, $\Delta E^{AH}_\pm = U_{\text{eff}}/2 + \alpha \omega_0 = U/2$, illustrated by the blue spectral peak where the most probable phonon excitation energy $\alpha \omega_0$ precisely cancels the polaronic reduction $-\alpha \omega_0$ (Fig. \ref{fig:bottomlineschematic}a). The spectral part of the impurity Green's function in the atomic limit \eqref{eq:atomicgreensfunc} and the functional envelope (Fig. \ref{fig:slowspectral}), calculated using the phonon probability distribution \eqref{eq:poisson}, confirm this compensation effect for the slow phonons. The central peak of the phonon excitation envelope, interpreted as the upper and lower Hubbard band positions, remains pinned at the same value $\pm U/2$ as in the bare single impurity Anderson model due to this subtle compensation and the impurity forms a local moment. Physically, this occurs because the slow phonons lack sufficient time to form the polaron and effectively screen the onsite interaction.

When the hybridization $V$ between the impurity and conduction bath is introduced from the atomic limit, the phonon peaks (Fig. \ref{fig:slowspectral}), which are separated by $\omega_0 \ll \Delta \ll U/2$ (where $\Delta$ is the resonance width from impurity conduction hybridization), merge to form complete upper and lower Hubbard bands centered around $\pm U/2$, identical to the single impurity Anderson model. Because electrons in the single impurity Anderson-Holstein model become agnostic to slow phonons, even with strong electron-phonon coupling, the low-energy physics remains identical to that of the single impurity Anderson model without phonons—manifesting as the Kondo effect where the phonons effectively do nothing. Virtual fluctuations between the low-energy spin manifold ($\vert f^1: \uparrow\rangle$,$\vert f^1: \downarrow\rangle$) and doubly occupied ($\vert f^2 \rangle$) or empty ($\vert f^0 \rangle$) impurity states with excitation energy $U/2$ lead to Kondo spin exchange and Kondo screening of the impurity spin by conduction electrons. The resulting low-energy effective theory is a spin Kondo model.

We now verify this physical intuition using perturbation theory for the full Anderson-Holstein Hamiltonian $H$ \eqref{eq:andersonholsteinHammy} in the shifted phonon basis $\tilde{H} = e^{i\hat{p}x_0} H e^{-i\hat{p}x_0}$,
\begin{equation}
\tilde{H} = \omega_0 \tilde{b}\dg \tilde{b} + \frac{U_{\text{eff}}}{2} (n_f - 1)^2 + V\sum_\sigma \left(c\dg_\sigma f_\sigma D\dg + \text{H.c.} \right),
\end{equation}
where the changes in the phonon cloud are incorporated in a redefinition of the hybridization. Since we assume that the single impurity Anderson model without phonons is in the spin Kondo phase ($\Delta \ll U$), we can also incorporate the virtual fluctuations using the Schrieffer-Wolff transformation (see Appendix \ref{sec:details} for details). 

The net amplitude for a  virtual valence fluctuation involves addition, followed by removal of an $f$ (or vice versa), as illustrated in Fig. \ref{fig:anisotropyflucs}a and Fig. \ref{fig:anisotropyflucs}b), given by  
\begin{equation}
\langle \tilde{0} \vert D\dg \vert \tilde{n}_b \rangle \langle \tilde{n}_b \vert D \vert \tilde{0} \rangle= \frac{\alpha^{\tilde{n}_b}}{\tilde{n}_b!}e^{-\alpha},
\end{equation}
using the matrix elements calculated in \eqref{eq:matrixelement}. The resulting amplitude is the sharply peaked Poisson distribution resembling a delta-function peaked at $\tilde{n}_b \approx \alpha$ excited phonons discussed earlier \eqref{eq:poisson}. When we sum over all $\tilde{n}_b$ we get a weight $1$, and since the excitation energy of the virtual fluctuation is $U/2$,  the Kondo spin exchange processes that  return to the same valence state have a coupling proportional to $J_K = 4V^2/U$. 

By contrast, the amplitude derived from a pair flip from $f^0$ to $f^2$ or vice versa, involves two $f$ additions or subtractions (see Fig. \ref{fig:anisotropyflucs}c) is then proportional to
\begin{equation}
\langle \tilde{0} \vert D \vert \tilde{n}_b \rangle \langle \tilde{n}_b\vert D \vert \tilde{0} \rangle = \frac{(-\alpha)^{\tilde{n}_b}}{\tilde{n}_b!}e^{-\alpha} 
\end{equation}
The magnitude of this quantity is also sharply peaked at $\tilde{n}_b \approx \alpha$, but summing over the $\tilde{n}_b$ to include all possible transitions, the result is an exponential suppression
$e^{-2\alpha}$, associated with the overlap between the polaron clouds of the $f^0$ and $f^2$ configurations, of the pair flip amplitude- otherwise known as the ``transverse'' fluctuations- with coupling proportional to $J_K e^{-2\alpha}$ because the excitation energy of the virtual fluctuation is also $U/2$.

Consequently, while the impurity spin degrees of freedom ($\vert f^1: \uparrow\rangle$ and $\vert f^1: \downarrow\rangle$) can undergo the Kondo effect where the phonon bath stays in the empty vacuum state $\vert 0_b \rangle$ because they are too slow to react, the impurity charge degrees of freedom ($\vert f^2 \rangle$ and $\vert f^0 \rangle$) cannot undergo Kondo charge exchange due to strong anisotropy in virtual valence fluctuations. Only ``Ising'' fluctuations that return the impurity to the same valence state remain unsuppressed, with ``transverse'' fluctuations exponentially suppressed. \lhl{In fact, the Kondo temperature for the anistropic Kondo model is exponentially suppressed compared to the isotropic Kondo model , $T_{K, \text{anisotropic}} \sim \left(T_{K, \text{isotropic}}\right)^{2\alpha}$ \cite{wiegmannandtsvelick} and is not accessible within our numerics}. We call this phase where virtual fluctuations keep the impurity valence unchanged from $f^0$ or $f^2$ in the \textit{frozen mixed valence} phase. We highlight that the frozen mixed valence phase is facilitated by a frozen polaron. For example, when the impurity virtually fluctuates from $f^0$ to $f^1$ and back, in the untranslated basis the phonons begin in the state $\vert + \rangle = e^{i px_0[n_f = 0]} \vert \tilde{0} \rangle$ and remain in that polaronic state during the ``Ising'' fluctuations because the phonons are too slow to react.

Their respective ground state energies determine the competition between the spin Kondo (SK) effect and the frozen mixed valence (FMV) phase. We predict a threshold dimensionless electron-phonon coupling $\alpha^*$ defined by, 
\begin{equation}
    U - 2\alpha^* \omega_0 = E_{SK} - E_{FMV}.
\end{equation}
Beyond this threshold $\alpha^*$, the physics is governed by the frozen mixed valence phase where the impurity valence is stuck in a state of $f^0$ or $f^2$. The spin Kondo effect's ground state energy is more negative than the frozen mixed valence phase due to the strong anisotropy of virtual fluctuations for the frozen mixed valence phase (Fig. \ref{fig:anisotropyflucs}b and Fig. \ref{fig:anisotropyflucs}c). Hence we predict $\alpha^* > \alpha_0$, where $\alpha_0 = U/(2\omega_0)$ is the dimensionless electron-phonon coupling that nullifies the effective onsite interaction $U_{\text{eff}} = U - 2 \alpha_0 \omega_0 = 0$.

\subsection{Dynamics in the Fast Phonon Regime}
In the fast regime, where the phonon frequency exceeds half the bare interaction strength ($\omega_0 \gtrsim U/2$), the system exhibits markedly different behavior from the slow phonon case. The phonons can respond dynamically to the changes in the impurity electronic occupation, enabling effective screening of the onsite Coulomb interaction. One way of understanding this is by looking at the phonon-mediated electron-electron interaction when the electron-phonon coupling is perturbative. Taking the phonon frequency to infinity in such a way as to keep $g^2/\omega_0 = \alpha \omega_0$ constant, the mediated interaction becomes instantaneous and attractive,
\begin{equation}
    \frac{\alpha \omega_0^2}{2} \frac{\omega_0} {\omega^2 - \omega_0^2} (n_f - 1)^2 \xrightarrow[]{\omega_0 \rightarrow \infty} -\frac{\alpha \omega_0}{2} (n_f - 1)^2, 
\end{equation}
which agrees with the polaronic reduction in \eqref{eq:completesquare}.

\begin{figure}[t!]
    \centering
    \includegraphics[width=0.9\linewidth]{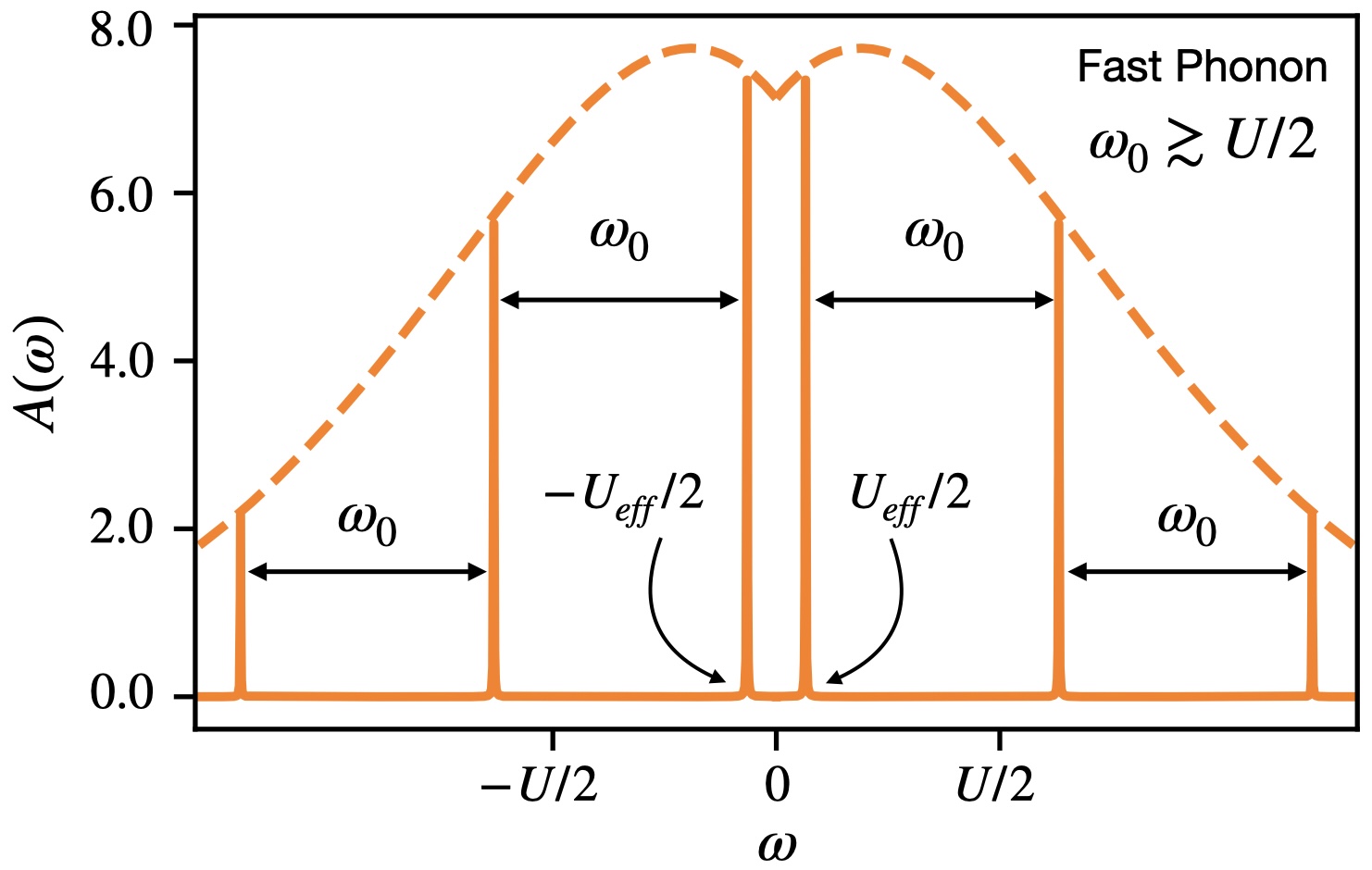}
    \caption{Impurity spectral function peaks and envelope function for the Anderson-Holstein model in the atomic limit, calculated using the Poisson distribution for the number of excited phonons ($\tilde{n}_b$). This figure shows the case where phonons are fast compared to the onsite interaction ($\omega_0 \gtrsim U/2$). The first peaks flanking zero energy are the renormalized interaction scale $\pm U_{\text{eff}}/2 = U/2 - \alpha \omega_0$, close to the peak of the envelope. The key energy scales are renormalized by coupling to a fast phonon mode because the phonon cloud can respond in time to charge fluctuations.}
    \label{fig:fastspectral}
\end{figure}

When the phonons are fast ($\omega_0 \gtrsim U/2$) and the dimensionless electron-phonon coupling is small ($\alpha \ll 1$), the Poisson distribution for phonon excitation probability Eq. \eqref{eq:poisson} peaks at $\tilde{n}_b = 0$. This results in a renormalized excitation energy for electron addition or removal from the singly occupied impurity: $U_{\text{eff}}/2 = U/2 - \alpha \omega_0$, without the compensation seen for the slow phonon case (Fig. \ref{fig:bottomlineschematic}c). This renormalization occurs because the polaron's phonon cloud can respond quickly enough to charge fluctuations to effectively screen the onsite interaction. Phonon side peaks with energy $\Delta E^{AH}_\pm = U_{\text{eff}} + \tilde{n}_b \omega_0$  persist even when the impurity-bath hybridization is introduced, as the phonon frequency substantially exceeds the hybridization width $\Delta$. The spectral part of the impurity Green's function in the atomic limit \eqref{eq:atomicgreensfunc} and the functional envelope (Fig. \ref{fig:fastspectral}), calculated using the phonon probability distribution \eqref{eq:poisson}, confirm this. When the hybridization $V$ between the impurity and conduction bath is introduced from the atomic limit, the lowest excitation peak (Fig. \ref{fig:fastspectral}) broadens to form upper and lower Hubbard bands centered around $\pm U_{\text{eff}}/2$ with additional separate phonon peaks with separation $\omega_0 \gg U/2 \gg \Delta$. 

To summarize, the fast phonons can react to the charge fluctuations, consequently renormalizing the key electronic energy scales of the single impurity Anderson model. The bare resonance width $\Delta$ is exponentially renormalized,
\begin{equation}
    \Delta \rightarrow \Delta \exp{\left[-\alpha \coth(\beta \omega_0 /2)\right]} \xrightarrow[]{T = 0} \Delta e^{-\alpha},
\end{equation}
because the adjustment of the bulky phonon cloud reduces the effectiveness of the tunneling of the dressed polaron into and out of the impurity. Simultaneously, the onsite interaction is also reduced by the polaron attraction,
\begin{equation}
    U \rightarrow U_{\text{eff}} = U - 2\alpha \omega_0.
\end{equation}
Moreover, phonon satellites arising from $\tilde{n}_b$ phonon excitations on top of the polaronic cloud appear in the local spectral function at distinct energies,
\begin{equation}
    \Delta E^{AH}_\pm = U_{\text{eff}} + \tilde{n}_b \omega_0.
\end{equation}
Ultimately, the coupling between a single impurity Anderson model to a fast phonon model renormalizes the single impurity physics at low energy and intermediate energy scales, adding additional phonon excitation features, as illustrated in Fig. \ref{fig:fastspectral}.

In contrast to the slow phonon regime, where a transition from spin Kondo to frozen mixed valence phases occurs above a threshold electron-phonon coupling $\alpha^*$, the fast phonon regime exhibits different behavior. When the electron-phonon coupling exceeds $\alpha_0 < \alpha^*$ (where the effective onsite interaction vanishes, $U_{\text{eff}} = U - 2\alpha_0 \omega_0 = 0$), the low-energy physics transitions from spin Kondo to charge Kondo effect.

We emphasize the fundamental difference between slow and fast phonon regimes. In the slow phonon regime, the low energy spin Kondo effect of the unrenormalized single impurity Anderson model is simply the conventional Kondo effect of bare electrons without phonon clouds. Conversely, in the fast phonon regime, polarons form first, followed by either spin Kondo or charge Kondo effects for the preformed polaron. 
 
\section{Numerical Renormalization Group Results}\label{sec:NRGsec}

We use the Numerical Renormalization Group~(NRG)~\cite{Wilson1975,Bulla2008,Weichselbaum2007,Weichselbaum2012} to verify our results from the previous sections. 
Our NRG calculations are done using the MuNRG package~\cite{Lee2016,Lee2017:PRL}, which is based on the QSpace tensor library~\cite{Weichselbaum2012a,Weichselbaum2020,Weichselbaum2024}. The data that support the findings of this article are openly available \cite{lau2025dataset}.
We use a hybridization function $\Delta(\omega)$ which has a box-shaped spectrum of half-width 1 (our unit of energy),
\begin{align}
-\tfrac{1}{\pi} \mathrm{Im} \, \Delta(\omega) = \tfrac{\Delta}{\pi} \theta(1-|\omega|) \, .
\end{align}
$\Delta(\omega)$ is discretized logarithmically with discretization parameter $\Lambda$.
The discretized impurity model is then mapped to a Wilson chain whose spectrum is determined via iterative diagonalization. 
We exploit $\mr{U}_c(1)\times \mr{SU}_s(2)$ charge and spin symmetries to reduce computational cost.
Explicit values for $\Lambda$, the number of kept multiplets, the considered size of the phonon Hilbert space, and our choice for $z$-averaging are parameter-dependent and are provided in the discussion below.

For our analysis, we compute impurity spectral functions for $f_{\sigma}$ and impurity contributions to thermodynamic properties, specifically the entropy and the spin and charge susceptibilities, denoted $S_{\mr{imp}}(T)$, $\chi_s(T)$ and $\chi_c(T)$, respectively.
For the spectral functions, we use the full density matrix NRG~\cite{Weichselbaum2007,Weichselbaum2012} together with the equations of motion trick of Ref.~\onlinecite{Kugler2022}. 
Thermodynamic properties are computed via the traditional approach as described, for instance, in Ref.~\onlinecite{Bulla2008}. 

\begin{figure}[t!]
\includegraphics[width = \linewidth]{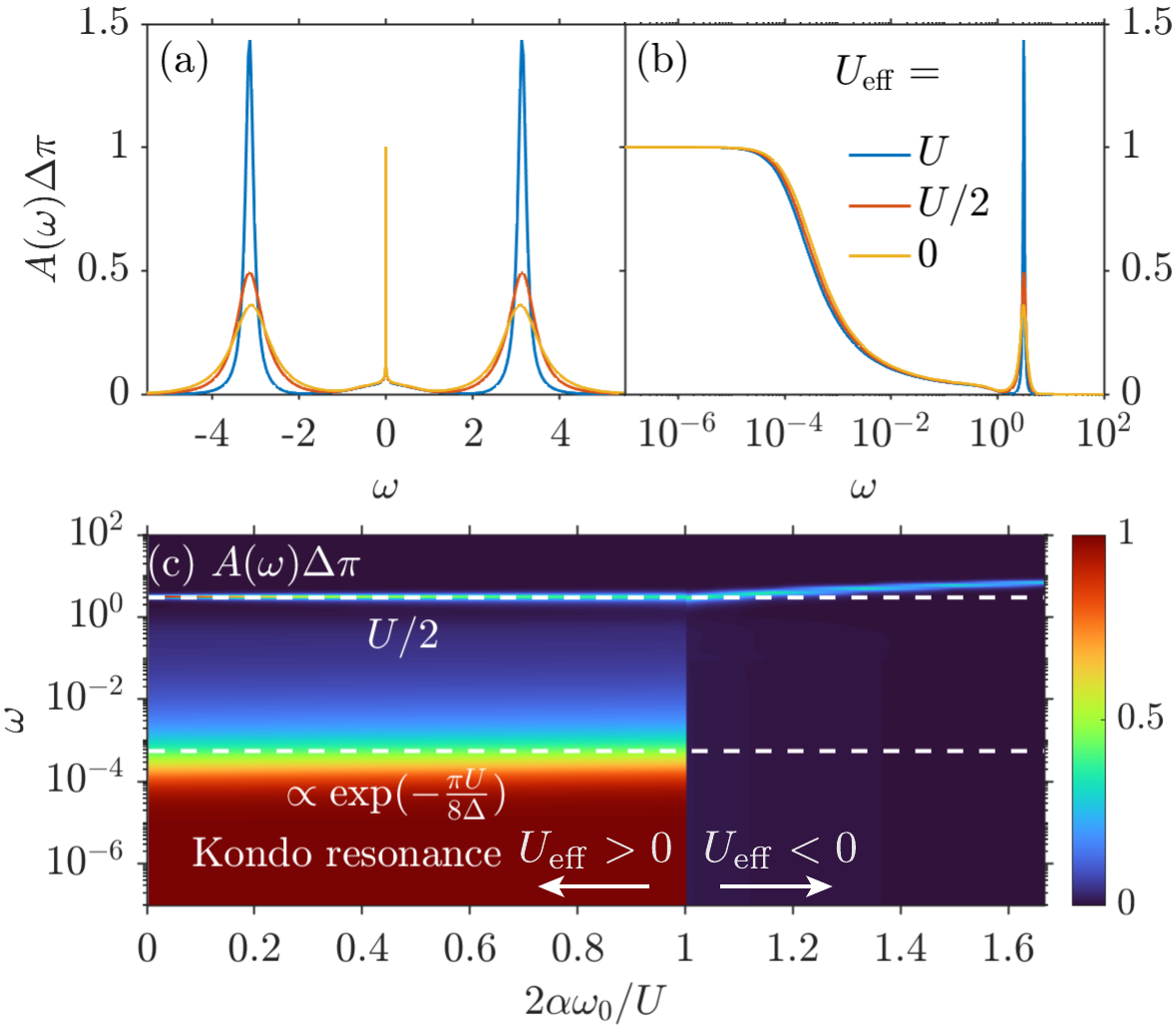}
\caption{
\label{fig:Spectra_w0small_U6_alphadep}
Spectral functions for an Anderson Holstein model with $U=6$ and $\Delta/\pi= \omega_0 = 0.1$, for different choices of $\alpha$.
(a,b) Impurity spectral function at selected values of $U_{\mr{eff}} = U - 2\alpha \omega_0$, on a linear and a logarithmic frequency scale,
(c) $\alpha$-dependence of the impurity spectral function, where the Kondo resonance disappears at a value $2\alpha^*\omega_0/U > 1$.
}
\end{figure}

\subsection{Slow phonon regime: $\omega_0 \ll U$}

We first consider a case where $\omega_0 = 0.1 \ll U = 6$ and $\Delta/\pi = 0.1$. We keep $\omega_0$, $U$ and $\Delta$ constant, while gradually increasing $\alpha$.
Due to the small phonon frequency, our phonon basis contains $2,000$ phonon number eigenstates to ensure converged results.
We further use $\Lambda = 2.5$ (which is relatively large for a single-orbital impurity model) and keep up to $4,000$ $\mr{U}_c(1)\times \mr{SU}_s(2)$ multiplets,
which ensures that our truncation energy is always $> 9$ in terms of rescaled energy units. We do not perform $z$-averaging in the small $\omega_0$ case.

In Fig.~\ref{fig:Spectra_w0small_U6_alphadep}(a,b), we show the spectral functions for $\alpha = 0$ ($U_{\mr{eff}} = U$), $\alpha = \alpha_0/2 = U/\omega_0$ ($U_{\mr{eff}} = U/2$),  and 
$\alpha = \alpha_0 = U/2\omega_0$ ($U_{\mr{eff}} = 0$), on a linear and logarithmic frequency scale, respectively.
At $\alpha = 0$ (no electron-phonon coupling), the spectral function $A(\omega)$ exhibits the expected features of a single-impurity Anderson model in the Kondo regime: two side peaks at $\omega = \pm U/2$ and a central Kondo resonance of width $T_K \propto \exp(-\pi U/8\Delta)$ and height $1/\Delta \pi$. 
These features remain remarkably unchanged as we turn on the electron-phonon coupling to $\alpha>0$. Even for a sizable electron-phonon coupling $\alpha = \alpha_0 = 30$, where $U_{\mr{eff}} = 0$, the electrons seem to be unaware of the presence of the phonons. 
The reason for that, as discussed above, is that the phonon is too slow to respond to the valence fluctuations which lead to the Kondo effect. Therefore, there is no significant renormalization of the Kondo resonance or the position of the
Hubbard side peaks, demonstrated here with an explicit numerical simulation.
\begin{figure}[t!]
\includegraphics[width = \linewidth]{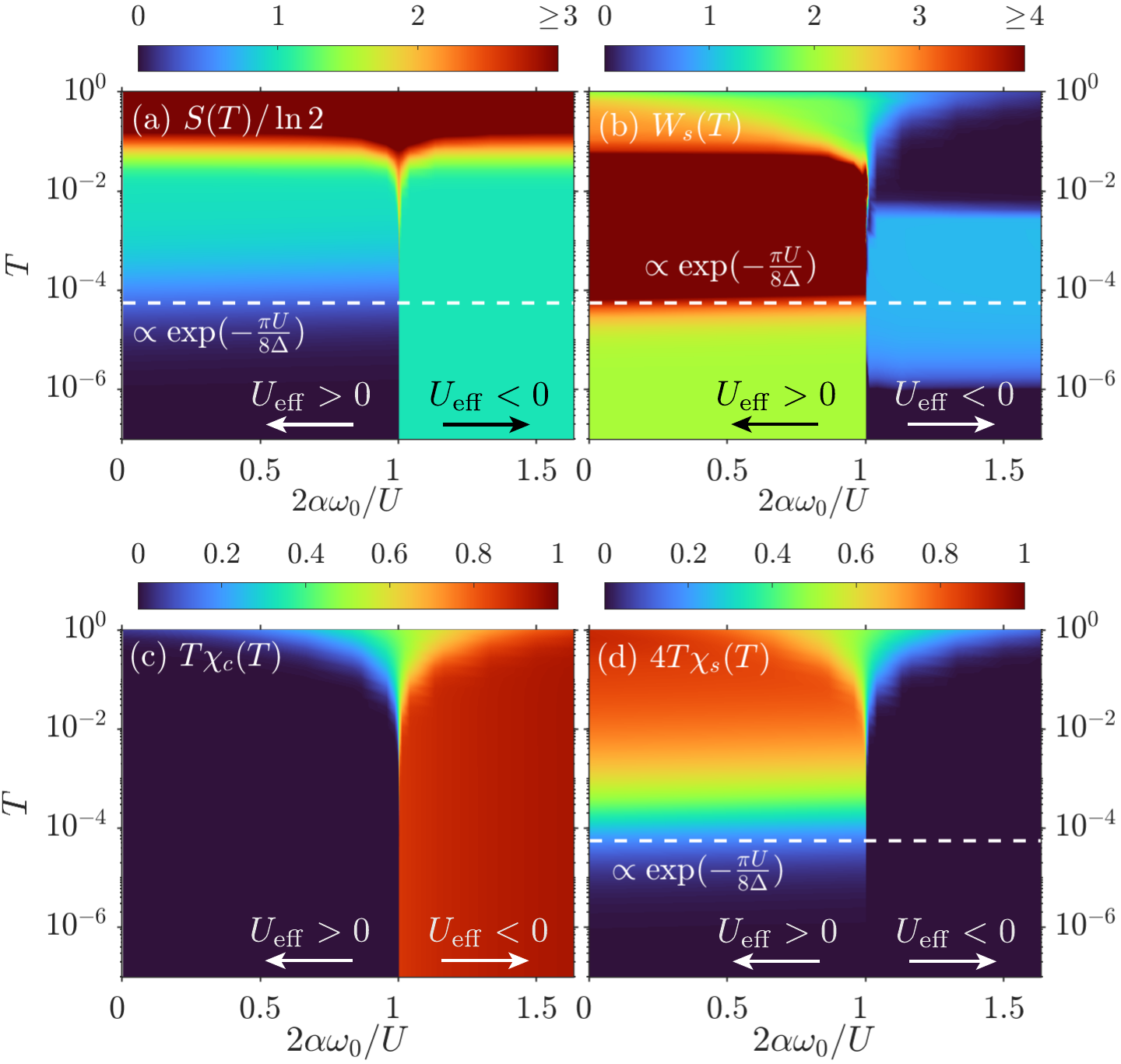}
\caption{
\label{fig:TD_w0_01_U6_alphadep}
Thermodynamic properties of an Anderson Holstein model with $U=6$ and $\omega_0 = \Delta/\pi = 0.1$, for different choices of $\alpha$.
(a) Impurity contribution to the entropy, (b) spin Wilson ratio, (c) impurity contribution to the charge susceptibility, and (d) impurity contribution to the spin susceptibility.
$U_{\mr{eff}} = U - 2\alpha\omega_0$.
}
\end{figure}

Figure~\ref{fig:Spectra_w0small_U6_alphadep}(c) shows the spectral function as a function of $\alpha$.
In the entire $\alpha < \alpha^{\ast}$ ($U_{\mr{eff}} > U^{\ast}_{\mr{eff}}$) region, the spectral function is remarkably $\alpha$-independent. 
It exhibits a Kondo resonance of width $T_{\mr{K}} \propto \exp(-\frac{\pi U}{8\Delta})$ and a side peak positioned at $\omega = \pm U/2$, i.e.\ all electronic energy scales depend 
only on the bare parameters $U$ and $\Delta$. 
The renormalized interaction $U_{\mr{eff}} = U - 2 \alpha \omega_0$ does not appear as a scale in the spectral function. 
The reason is that $U_{\mr{eff}}$ sets the energy difference 
between states with different phonon condensates, 
and the phonon condensate remains fixed after acting with $f_{\sigma}$ on the ground state sector. 
Since the phonon is then way too slow to adjust its condensate on electronic time scales, no renormalization effects occur. 

\begin{figure}[b!]
\includegraphics[width = \linewidth]{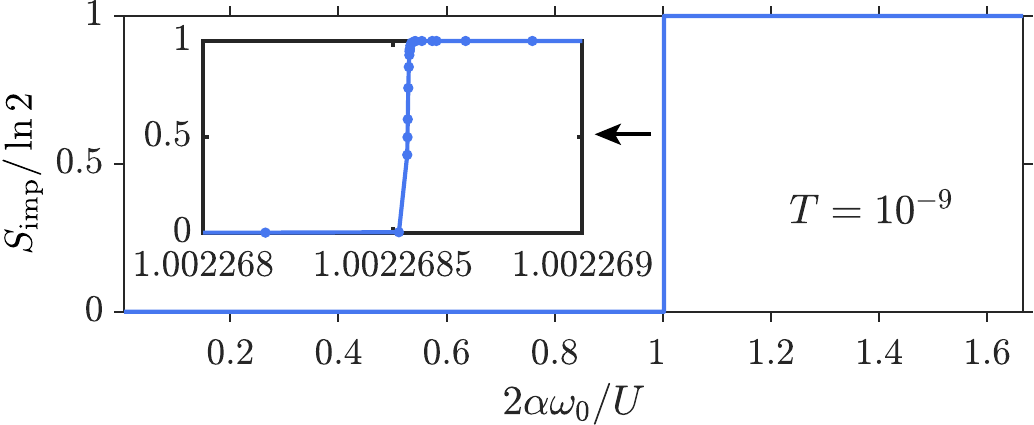}
\caption{
\label{fig:slowPhonon_Sent_vs_alpha}
Low-temperature ($T=10^{-9}$) impurity contribution to the entropy versus $\alpha$ in the slow phonon regime ($U = 6$, $\omega_0 = \Delta/\pi = 0.1$). Inset: Close-up of the crossover region at 
$2\alpha^{\ast}\omega_0/U = \alpha^{\ast}/\alpha_0 \simeq 1.0022685$.
Dots in the inset mark data points.
}
\end{figure}

For $\alpha > \alpha^{\ast}$ ($U_{\mr{eff}} < U^{\ast}_{\mr{eff}}$), the doublon and holon states with shifted phonon condensates are the ground state sector on the impurity.
The resulting charge doublet is extremely long-lived and does not undergo a charge Kondo effect at any reasonable temperature or time scale,
leading to a sudden disappearance of the Kondo resonance when crossing from $\alpha < \alpha^{\ast}$ ($U_{\mr{eff}} > U^{\ast}_{\mr{eff}}$) to $\alpha > \alpha^{\ast}$ ($U_{\mr{eff}} < U^{\ast}_{\mr{eff}}$), see Fig.~\ref{fig:Spectra_w0small_U6_alphadep}(c).
The crossover value $\alpha^{\ast}$ is slightly larger than $\alpha_0$ (where $U_{\mr{eff}} = 0$) due to the presence of virtual spin fluctuations in the local moment sector; see Fig.~\ref{fig:slowPhonon_Sent_vs_alpha} and its discussion below.
These lead to a further dynamical reduction of the ground state energy of the former with respect to the doublon/holon sector. 
Further, the disappearance of the Kondo resonance appears to be a very rapid crossover, not a (first-order) quantum phase transition.

In Fig.~\ref{fig:TD_w0_01_U6_alphadep}, we show that the phonons also do not considerably influence the thermodynamics when $U_{\mr{eff}} \geq 0$. 
The impurity contribution to the entropy [Fig.~\ref{fig:TD_w0_01_U6_alphadep}(a)] is $\sim \ln 2$ in the local moment
regime ($T > T_{\mr{K}}$) and subsequently reduced to 0 when the moment is screened below $T_{\mr{K}}$, except for a large phonon contribution at $T > \omega_0 = 0.1$. 
The spin Wilson ratio $W_s(T)$ is shown in Fig.~\ref{fig:TD_w0_01_U6_alphadep}(b). $W_s(T)$ is enhanced in the local moment region at $T > T_{\mr{K}}$ and finally reduced to $W_s(T) = 2$ in the Fermi liquid at $T < T_{\mr{K}}$, 
confirming that the impurity model undergoes a spin Kondo effect. This is also consistent with the temperature dependence of $T \chi_s(T)$ [Fig.~\ref{fig:TD_w0_01_U6_alphadep}(d)], which is constant in the local moment region (i.e., $\chi_s(T) \sim T^{-1}$ has a Curie susceptibility), and $T \chi_s(T)$ is subsequently reduced to 0 below $T_{\mr{K}}$ since the local moment is screened there. 
The charge susceptibility in Fig.~\ref{fig:TD_w0_01_U6_alphadep}(c), $T \chi_c(T)$ is strongly suppressed for $U_{\mr{eff}} \geq 0$, as expected. 

\begin{figure}[t!]
\includegraphics[width = \linewidth]{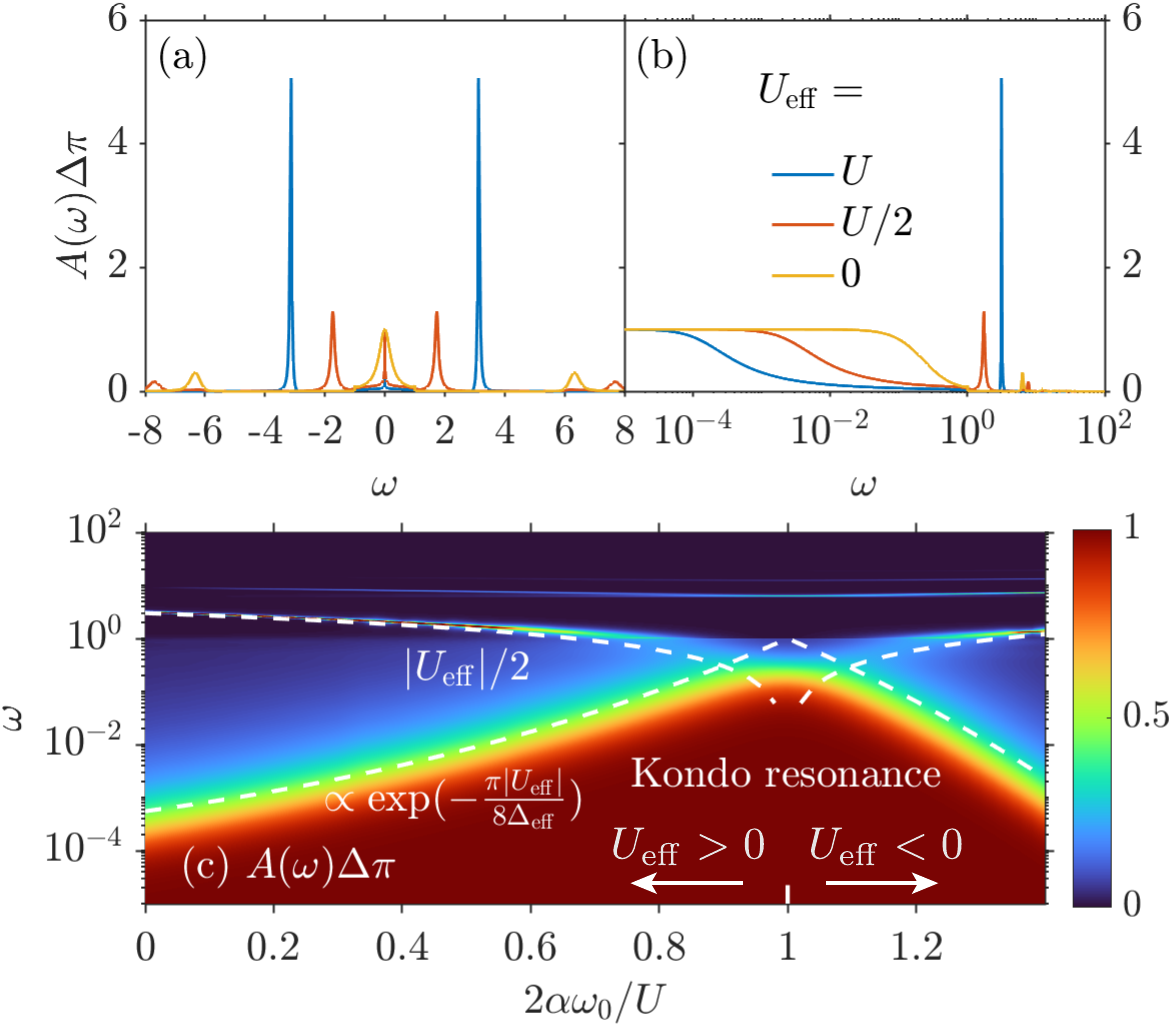}
\caption{
\label{fig:Spectra_w0_6_U6_alphadep}
Spectral functions for an Anderson Holstein model with $U=\omega_0=6$ and $\Delta/\pi = 0.1$, for different choices of $\alpha$.
(a,b) Impurity spectral function at selected values of $U_{\mr{eff}} = U - 2\alpha \omega_0$, on a linear and a logarithmic frequency scale and
(c) $\alpha$-dependence of the impurity spectral function.
$\Delta_{\mr{eff}} = e^{-\alpha} \Delta$, $U_{\mr{eff}} = U - 2\alpha\omega_0$.
}
\end{figure}

The thermodynamics of the $U_{\mr{eff}} < 0 $ region, on the other hand, is consistent with the formation of a charge doublet that does not undergo screening, even at the lowest temperatures considered here.
The impurity contribution to the entropy is $\sim \ln 2$ (except for $T > \omega_0$), the charge susceptibility has a Curie form, $T \chi_c(T) = \mr{const}$, and the spin susceptibility is strongly suppressed. 
Interestingly, the spin Wilson ratio $W_s(T) \simeq 1$ at intermediate to low temperatures, indicates th/at there are still thermodynamically active spin fluctuations above the doublon/holon ground state.
We note here that for $T < 10^{-5}$ and $U_{\mr{eff}} < 0$, our result for $W_s(T)$ is numerically not reliable because both the spin susceptibility and the specific heat become very small there.  $W_s$ is their ratio and thus very prone to numerical errors if $T \chi_s(T)$ and $C(T)$ are tiny. 
It is therefore not clear from our numerics whether $W_s(T) \simeq 1$ holds at even lower temperatures than indicated by our data in the $U_{\mr{eff}} < 0$ region. \lhl{For the anisotropic Kondo model when $U_{\text{eff}} <0$, we can estimate the Kondo temperature $T_{K, \text{anisotropic}} \sim \left(T_{K, \text{isotropic}}\right)^{2\alpha}$ \cite{wiegmannandtsvelick}. With $T_{K, \text{isotropic}} \sim 10^{-3}$ from Fig. \ref{fig:Spectra_w0small_U6_alphadep} and $2\alpha \omega_0/U \sim 1$, we obtain $T_{K, \text{anisotropic}} \sim 10^{-180}$, indicating our simulations operate well above the Kondo temperature in this regime. Thus, our calculations have only begun flowing from the weak-coupling fixed point and remain far (approximately 170 energy decades) from approaching the strong-coupling fixed point. This limitation prevents us from calculating the contributions of the two leading irrelevant operators, near the strong-coupling fixed point, to the susceptibility and specific heat, as accomplished by Wilson \cite{Wilson1975}.}

In the close vicinity of the crossover region around $\alpha^{\ast}$ ($U_{\mr{eff}} \simeq 0$), the behavior somewhat deviates from the two scenarios described above. 
For instance, we find that the large phonon-related entropy extends to temperatures considerably lower than $\omega_0$. However, since these deviations only occur 
in a very narrow region (and therefore require a lot of fine-tuning), and since these deviations further do not extend down to $T=0$, 
their relevance to realistic situations is quite limited and we therefore refrain from a closer investigation.

To showcase the low-temperature crossover between spin Kondo and frozen charge doublet, Fig.~\ref{fig:slowPhonon_Sent_vs_alpha} shows the impurity contribution to the entropy at $T=10^{-9}$ versus $\alpha$. Close to $\alpha = \alpha_0 = 30$, there is an almost step-like transition from $S_{\mr{imp}} = 0$ for $\alpha \ll \alpha_0$ to $S_{\mr{imp}} = \ln 2$ for $\alpha \gg \alpha_0$. 
The inset in Fig.~\ref{fig:slowPhonon_Sent_vs_alpha}
shows a close-up view of the crossover region, where the entropy continuously but rapidly transitions between the aforementioned limits. The inset also demonstrates that this crossover happens around $\alpha^{\ast} > \alpha_0$, due to virtual fluctuations in the spin doublet sector.

\begin{figure}[t!]
\includegraphics[width = \linewidth]{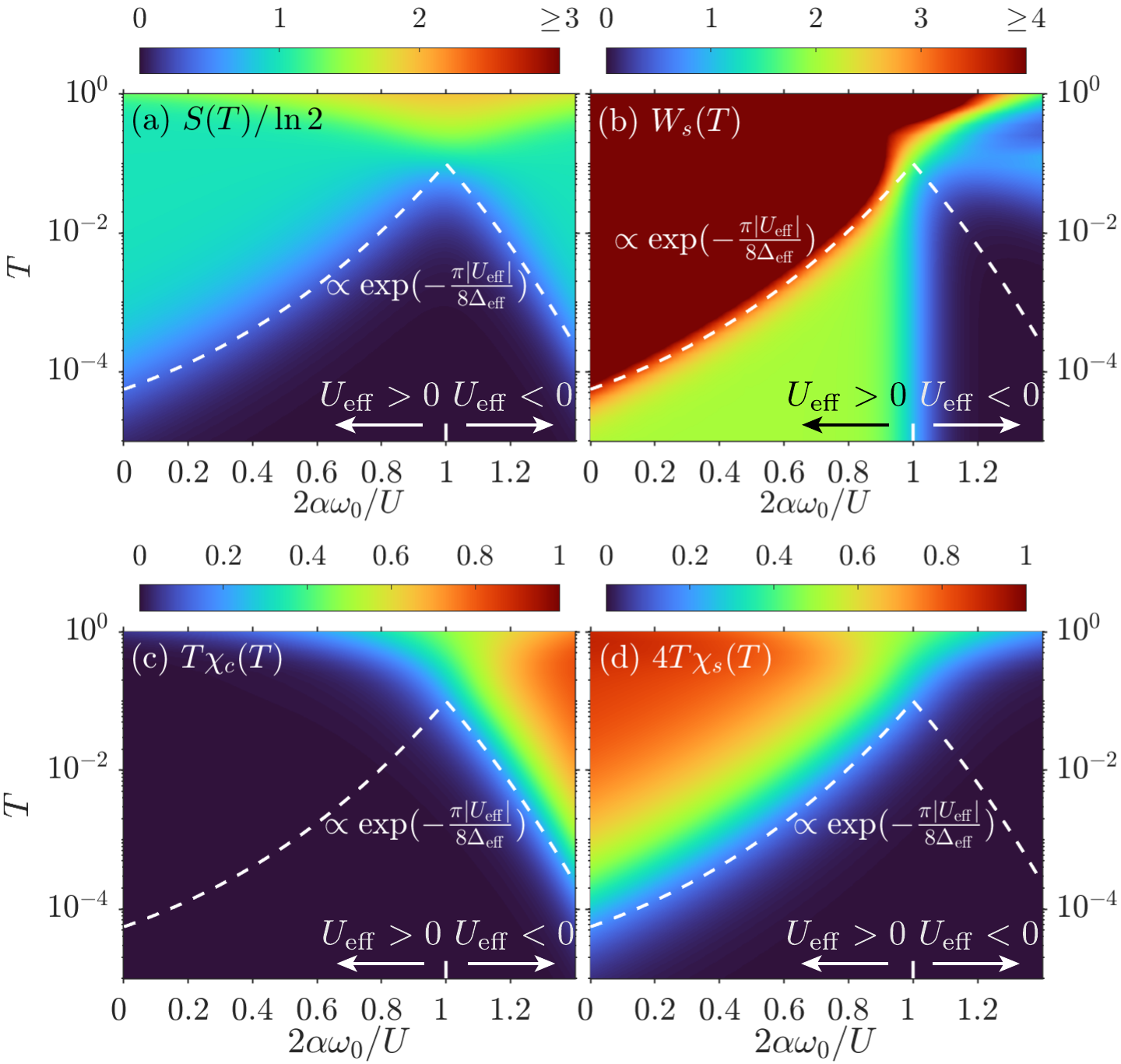}
\caption{
\label{fig:TD_w0_6_U6_alphadep}
Thermodynamic properties of an Anderson Holstein model with $U=\omega_0=6$ and $\Delta/\pi = 0.1$, for different choices of $\alpha$.
(a) Impurity contribution to the entropy, (b) spin Wilson ratio, (c) impurity contribution to the charge susceptibility, and (d) impurity contribution to the spin susceptibility.
$\Delta_{\mr{eff}} = e^{-\alpha} \Delta$, $U_{\mr{eff}} = U - 2\alpha\omega_0$.
}
\end{figure}

\subsection{Fast phonon regime: $\omega_0 \gtrsim U/2$}

A very different picture unfolds when phonons are fast enough to respond on electronic time scales. 
To study that with NRG, we consider the case where $U=\omega_0=6$, $\Delta/\pi = 0.1$, and tune $\alpha$.
We keep up to 100 phonon eigenmodes in our phonon basis, use $\Lambda = 2$ for logarithmic discretization,
average over two $z$-shifted discretizations, and keep up to $2,000$ $\mr{U}_c(1)\times \mr{SU}_s(2)$ multiplets, which ensures a truncation energy $> 12$ in terms of rescaled units.

Figure~\ref{fig:Spectra_w0_6_U6_alphadep}(a,b) shows the impurity spectral function for different values of $U_{\mr{eff}}$. 
In stark contrast to the slow phonon case [Fig.~\ref{fig:Spectra_w0small_U6_alphadep}(a,b)], the electron-phonon coupling now
clearly renormalizes the electronic spectral function: as $U_{\mr{eff}} = U - 2\alpha \omega_0$ is reduced (corresponding to increasing electron-phonon coupling $g = \sqrt{\alpha} \omega_0$),
the width of the central Kondo resonance increases by several orders of magnitude while the side peaks move toward the central resonance. 
At the same time, replica side peaks emerge at $\omega \simeq |U_{\mr{eff}}|/2 + n \omega_0$, with $n = 1,2,\dots$.
Their spectral weights decay exponentially with increasing $n$, see Fig.~\ref{fig:fastspectral} and its discussion. For that reason, only the first two replicas are clearly visible in Fig.~\ref{fig:Spectra_w0_6_U6_alphadep}(c),
where we show the spectral function with more fine-grained $\alpha$-dependence and for a larger frequency range.

\begin{figure}[t!]
\includegraphics[width = \linewidth]{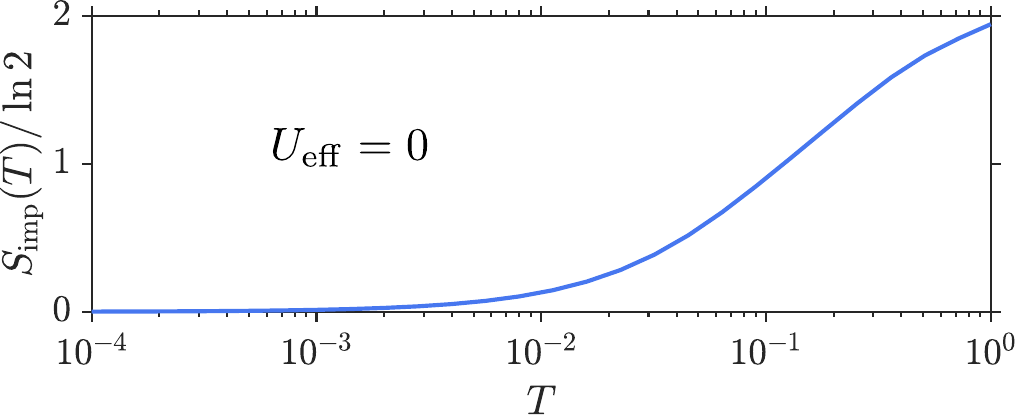}
\caption{
\label{fig:Ueff0_Simp}
\ag{Line cut of Fig.~\ref{fig:TD_w0_6_U6_alphadep}(a) at $2\alpha \omega_0/U = 1$ ($U_{\mr{eff}} = 0$).}}
\end{figure}

The data in Fig.~\ref{fig:Spectra_w0_6_U6_alphadep}(c) demonstrates that $A(\omega)$ 
is governed by the phonon-renormalized interaction $U_{\mr{eff}}$ and hybridization $\Delta_{\mr{eff}} = \Delta e^{-\alpha}$.
The width of the central resonance is given by $T_{\mr{K}}^{\ast} \propto \exp(-\frac{\pi |U_{\mr{eff}}|}{8\Delta_{\mr{eff}}})$, and the side peaks are positioned at $\omega = \pm |U_{\mr{eff}}|/2$. 
These relations also hold for $U_{\mr{eff}} < 0$ case, where the impurity model undergoes a charge Kondo effect. 
Further, there is a smooth crossover from a spin Kondo ($U_{\mr{eff}} > 0$) to a charge Kondo ($U_{\mr{eff}} < 0$) effect for fast phonons, 
as opposed to slow phonons, where we found an almost step-like crossover from a spin Kondo effect to an unscreened charge doublet.

\begin{figure}[b!]
\includegraphics[width = \linewidth]{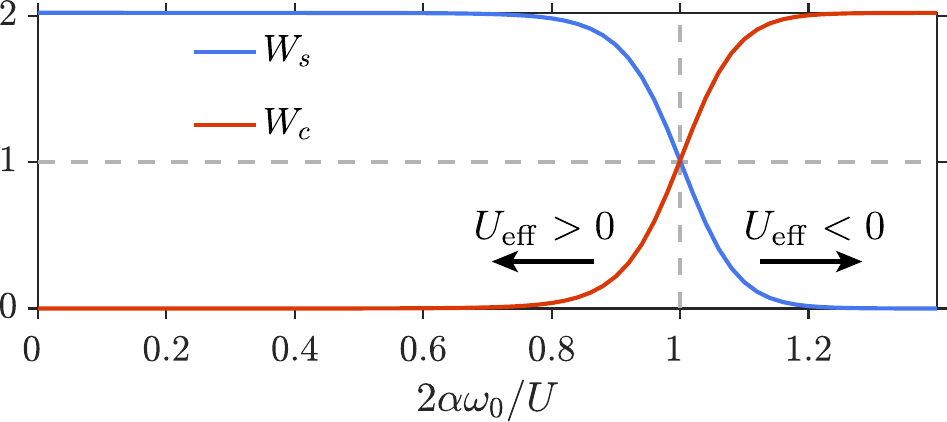}
\caption{
\label{fig:WilsonRatio_fastPhonon}
Crossover behavior of the low-temperature ($T \ll T_{\mr{K}}$) spin (blue line) and charge (red line) Wilson ratios versus $\alpha$ in the fast-phonon regime ($\omega_0 = U = 6$ and $\Delta/\pi = 0.1$). 
}
\end{figure}

The renormalization of the bare interaction and hybridization strength also influences the thermodynamic properties accordingly, shown in Fig.~\ref{fig:TD_w0_6_U6_alphadep}.
The impurity contribution to the entropy, Fig.~\ref{fig:TD_w0_6_U6_alphadep}(a), is $\sim \ln 2$ for $T > T_{\mr{K}}^{\ast}$ due to either a thermally fluctuating spin ($U_{\mr{eff}} > 0$)
or charge ($U_{\mr{eff}} < 0$) doublet. This doublet is screened below $T_{\mr{K}}^{\ast}$, where the impurity entropy is linear-in-$T$.
Due to the large phonon frequency, there is no phonon contribution to the entropy at the temperatures considered here. 
The fast phonon is, therefore, only dynamically active in the sense that it strongly renormalizes the electronic structure, but it is not thermodynamically active.
As discussed above, the converse is true for the slow phonon: it is thermodynamically active but not dynamically.

\ag{At $U = 2\alpha \omega_0$ ($U_{\mr{eff}} = 0$), no spin or isospin local moment is formed. The impurity contribution to the entropy (see Fig.~\ref{fig:Ueff0_Simp} for a line cut) decreases from $2 \ln 2$ to zero as the temperature is lowered (assuming $T \ll \omega_0$). Notably, there is no feature at $\ln 2$ that would indicate the formation of a local moment}

Figure~\ref{fig:TD_w0_6_U6_alphadep}(b) shows the Wilson ratio $W_s(T)$. For $U_{\mr{eff}} > 0$, $W_s(T)$ is strongly enhanced in the (magnetic) local moment
region at $T > T_{\mr{K}}^{\ast}$ and reduced to $W_s(T) \simeq 2$ blow $T_{\mr{K}}^{\ast}$ where the local moment is screened. 
At low temperatures $T < T_{\mr{K}}^{\ast}$ and as a function of $\alpha$, the Wilson ratio crosses over from $W_s \simeq 2$ at small $\alpha$ where $U_{\mr{eff}} > 0$, to $W_s \simeq 0$ at large $\alpha$ where $U_{\mr{eff}} < 0$.
This is consistent with a crossover from a spin Kondo effect to a charge Kondo effect, see also Fig.~\ref{fig:WilsonRatio_fastPhonon} and its discussion.
The crossover from spin to charge Kondo can also be observed in the impurity contribution to the charge and spin susceptibilities, see Fig.~\ref{fig:TD_w0_6_U6_alphadep}(c) and (d), respectively. 
For $U_{\mr{eff}} > 0$ and $T > T_{\mr{K}}^{\ast}$, $T \chi_s(T) \sim \mr{const}$ has a Curie form, while $T \chi_c(T)$ is small, and vice versa for $U_{\mr{eff}} < 0$.
This further demonstrates the expected presence of a spin doublet for $U_{\mr{eff}} > 0$ and of a charge doublet for $U_{\mr{eff}} < 0$ above the Kondo scale.
Below $T_{\mr{K}}^{\ast}$, the doublet gets screened and both $T \chi_s(T) \sim T$ and $T \chi_c(T) \sim T$.

Figure~\ref{fig:WilsonRatio_fastPhonon} demonstrates the crossover from spin to charge Kondo physics in terms of the $\alpha$-dependence of the low temperature ($T \ll T_{\mr{K}}^{\ast}$) spin and charge Wilson ratios.
Deep in the spin Kondo regime where $U \gg 2\alpha \omega_0$ ($U_{\mr{eff}} > 0$), the impurity model undergoes a pure spin Kondo effect. As a result, $W_s = 2$ while its charge counterpart vanishes, $W_c = 0$. 
Well within the charge Kondo regime where $U \ll 2\alpha \omega_0$ ($U_{\mr{eff}} < 0$), on the other hand, we find a pure charge Kondo effect, resulting in $W_s = 0$ and  $W_c = 2$.
The two regimes are connected by a smooth crossover. At $\alpha = 0.5$, where $U_{\mr{eff}} = 0$, $W_s = W_c = 1$, i.e.\ the spin and charge Wilson ratios are the same as is the case for a non-interacting resonant level.

\section{Discussion and Outlook}\label{sec:discussion}
Our study reveals a fundamental dichotomy in impurity Anderson-Holstein models governed by phonon dynamics. When phonons are fast relative to charge fluctuations ($\omega_0 \gtrsim U/2$), they rapidly form a polaronic cloud that screens the impurity, renormalizing the onsite interaction $U_{\text{eff}} = U - 2\alpha \omega_0$. Spectral satellites form from phonon excitations due to their ability to respond to charge fluctuations while the screening mechanism dramatically transforms the low-energy physics: as the effective interaction transitions from repulsive to attractive, the system crosses over between spin Kondo, mixed valence, and charge Kondo phases. Conversely, when the phonons are slow compared to charge fluctuations ($\omega_0 \ll U/2$), their dynamics become frozen, preventing the formation of a responsive polaronic cloud to screen the impurity. When an electron is added or removed,  a compensation develops between the  occupation energy of the excited-state phonons and the polaronic energy level, which  maintains the excitation energy at $U/2$, characteristic of the uncoupled single impurity Anderson model. Consequently, despite strong coupling,  slow phonons are inconsequential, leaving the electronic interactions of the single impurity Anderson model unchanged. Beyond a threshold electron-phonon coupling strength in the slow phonon regime, we predict a transition from the usual spin Kondo effect to a frozen mixed valence phase where the static phonon cloud around the impurity locks the system into either $f^0$ or $f^2$ valence configurations, with ``Ising'' virtual fluctuations which return to the same valence state. \lhl{We note that the transition from a local moment to charge doublet picture becomes extremely sharp when the phonon frequency $\omega_0$ approaches zero while electron-phonon coupling remains finite. This occurs because the charge and spin sectors decouple completely, causing the transition to resemble a first-order transition at zero temperature. }

\lhl{When tuning from the slow phonon regime (Figs. \ref{fig:Spectra_w0small_U6_alphadep}-\ref{fig:TD_w0_01_U6_alphadep}) to the fast phonon regime (Figs. \ref{fig:Spectra_w0_6_U6_alphadep}-\ref{fig:TD_w0_6_U6_alphadep}), we expect a gradual transition at $\omega_0 = U/2$ because both phases exhibit a scattering phase shift of $\pi/2$ at temperatures below the Kondo temperature. This applies even in the frozen mixed valence phase (negative $U_{\text{eff}}$ phase in Fig. 7(c)), where the Kondo temperature is incredibly suppressed due to the strong anisotropy of the Kondo coupling.}

While our study has focused on the symmetric Anderson-Holstein impurity model, further work is needed to extend it to the asymmetric case, though we believe the qualitative physics would still hold. In the asymmetric case, the frozen mixed valence phase will develop between impurity valences that differ by one instead of two (as in the symmetric case). The  coexistence of static mixed valence behavior ($\textrm{Fe}^{2+}$ and $\textrm{Fe}^{3+}$) with insulating properties in rust is likely  related to the frozen mixed valence phase of the asymmetric Anderson-Holstein model. This connection is strengthened by observations of reduced and structurally disordered metal sites in rust, consistent with small polaron formation where charge carriers become trapped through strong electron-lattice interactions \cite{rustsmallpolaron}. Future investigations could explore whether phonon timescales in iron oxides facilitate this frozen polaron mechanism, potentially explaining the mystery of insulating behavior in mixed valence rust systems.

{The role of the ratio of phonon frequency to onsite $U/2$ in interaction renormalization is not limited to Anderson-Holstein  models. In a recent determinant quantum Monte Carlo study \cite{zhuholstein2025} to systematically explore the potential of d-wave superconductivity in the Hubbard-Holstein model, polaron driven renormalizations of the onsite Hubbard interaction, that appeared only in the fast phonon regime, were also numerically obtained.}

Our findings provide a promising perspective for materials where the phonon scale is comparable to or larger than the Coulomb scale. A notable example is twisted bilayer graphene (TBG), where recent experiments ~\cite{chen_strong_2024, birkbeck_measuring_2024} have revealed strong electron-phonon coupling between flat bands and phonons with frequencies $\omega_0 \sim 150 \, \textrm{meV}$. Discrepancies between theoretical predictions of the onsite interaction ($U_{\text{theory}} \sim 60 - 100 \, \textrm{meV}$ \cite{song21, Lau23, lau2025}) and experimental observations from scanning tunneling and quantum twisting microscopy \cite{qtm2023} ($U_{\text{expt}} \sim 23-35 \, \textrm{meV}$ \cite{cascade, qtm2025}) have been highlighted by Lau and Coleman \cite{lau2025}. This suggests that strongly coupled phonons may operate on faster timescales than charge fluctuations, potentially leading to polaronic screening and renormalization of the onsite $U$. This direction is particularly intriguing, as theoretical work has proposed that one of the fast phonon modes ($K$-phonon) could drive superconductivity through a phonon-mediated valley Hund’s interaction \cite{zhida1,zhida2}.

Additionally, the alignment of hexagonal boron nitride (hBN) to TBG may play a crucial role in the electron-phonon coupling dynamics. When TBG is misaligned with hBN, optical graphene phonons oscillate freely at approximately $150\, \textrm{meV}$, manifesting as prominent replica bands in spectroscopic measurements \cite{chen_strong_2024}, with considerable electron-phonon coupling strength \cite{birkbeck_measuring_2024}. However, when TBG aligns with hBN, we expect the collective atomic motion to stiffen the graphene phonons,  increasing the phonon frequency $\omega_0$ beyond typical measurement ranges while affecting the e-ph coupling in a non-trivial way~\cite{jung2015vibrational}.
This stiffening would shift replica bands to higher energies while simultaneously reducing the dimensionless phonon coupling $\alpha$ for fixed coupling strength $g = \sqrt{\alpha} \omega_0$- diminishing the spectral weight of these phonon satellites in hBN aligned samples. We note that superconductivity also disappears in these aligned samples, raising the intriguing question of whether the substrate-induced phonon stiffening directly impacts the superconducting mechanism in TBG.

Future research could investigate how the low-energy phases of the Anderson-Holstein model change when the conduction bath has a linear Dirac spectrum, rather than parabolic, band structure, as explored in this study. For instance, in a linear density of states, it is known that the single impurity Anderson model exhibits an unstable fixed point and a modified strong coupling regime where the local moment remains partially unscreened \cite{Withoff1990}. 

The role of electron-phonon coupling in strange metals represents another compelling area for study, particularly near quantum critical points where phonons may transition from being slow to being fast relative to electronic degrees of freedom. Building on observations in $\textrm{YbBAl}_4$ \cite{kobayashi23}, where electrons become slow degrees of freedom near quantum criticality and exhibit polaronic response, we propose extending our framework to models of heavy fermion quantum criticality \cite{andreasprx, andreasprl} to better understand strange metal behavior. The effect of electron-phonon coupling on $f$-level magnetism and associated symmetry-breaking \cite{huang2024hidden,kushnirenko2024unexpected} is a promising future application of this work.
This approach may reveal how the phonon timescale crossover impacts the non-Fermi liquid physics and unusual transport properties characteristic of these systems.

\begin{acknowledgements}
The authors are grateful for important conversations with X. Dai, J. Huang, S. Ilani, E. Khalaf, and C-X Liu.
This work was carried out in part at the Aspen Center for Physics (D.K., Pr.C., Pi.C), which is supported by the National Science Foundation grant PHY-2210452.
Additionally, D.K. acknowledges partial support by grant NSF PHY-2309135 to the Kavli Institute for Theoretical Physics (KITP), where part of this work was performed. A.G. and D.K. are supported by the Abrahams Postdoctoral fellowship of the Center for Materials Theory, Rutgers University. D.K. acknowledges support from the Zuckerman STEM Fellowship. L.L.H.L. acknowledges support from the Lindenfeld Graduate Fellowship from Rutgers University.
This work was supported by the Office of Basic Energy Sciences, Material Sciences and Engineering Division, U.S. Department of Energy (DOE) under contract DE-FG02- 99ER45790 (L.L.H.L. and Pi.C.).
\end{acknowledgements}
 


\bibliography{bibliography,combined7}
\appendix
\begin{widetext}
\section{Lang-Firsov Transformation}\label{sec:langfirsovappendix}
Starting from the atomic limit of the impurity Anderson-Holstein model,
\begin{eqnarray}\label{eq:atomicAHappendix}
    H_{\text{atomic}} &=& \omega_0 b\dg b + \sqrt{\alpha} \left(b + b\dg\right)(n_f - 1) + \frac{U}{2} (n_f-1)^2 \cr &=& \omega_0 (b\dg+ \sqrt{\alpha} (n_f-1))( b + \sqrt{\alpha} (n_f-1)) +\frac{\left(U - 2\alpha \omega_0 \right)}{2} (n_f-1)^2,
\end{eqnarray}
where $b\dg$ creates the Einstein phonon with frequency $\omega_0$, $n_f = \sum_\sigma f\dg_\sigma f_\sigma$ is the number operator of the impurity electrons, and $\sqrt{\alpha} \omega_0$ is the electron-phonon coupling. The phonon oscillator position and momentum operators are $x = (b + b\dg)/\sqrt{2  \omega_0}$ and $p = i\sqrt{(\omega_0)/2}(b\dg - b) $ respectively. We rewrite the Hamiltonian \eqref{eq:atomicAHappendix} as a displaced harmonic oscillator,
\begin{equation}
    H_{\text{atomic}} = \frac{\hat{p}^2}{2} + \frac{ \omega_0^2}{2}(\hat{x}-x_0[n_f])^2 + \frac{(U - 2\alpha \omega_0)}{2}(n_f - 1)^2,
\end{equation}
where the linear electron-phonon coupling term has been eliminated and the sign of the displaced equilibrium of the phonon oscillator
\begin{equation}\label{eq:displacedeqm}
   x_0[n_f] = \sqrt{\frac{2\alpha}{\omega_0}} (1- n_f )
\end{equation}
depends on whether the impurity is empty or doubly occupied. 

We can perform the Lang-Firsov transformation which is a unitary transformation that translates the phonon oscillator reference position to its new equilibrium position $x_0$ by the translation operator $e^{-ipx_0}$ which is defined as,
\begin{equation}
    e^{-i\hat{p}x_0} = \exp{\left[\sqrt{\alpha}\left(b - b\dg\right)(n_f - 1)\right]},
\end{equation}
which acts on wavefunctions to translate the position by $x_0$, $e^{ipx_0} \psi(x) = \psi(x+x_0)$. Acting the unitary transformation on $H_{\text{atomic}}$, 
\begin{eqnarray}\label{eq:displacedHammy}
    e^{i\hat{p}x_0} H_{\text{atomic}} e^{-i\hat{p}x_0} = \tilde{H} &=& \sum_{\bk \sigma} \epsilon_{\bk} c\dg_{\bk \sigma} c_{\bk \sigma} + \omega_0(\tilde{b}\dg \tilde{b} +1/2) +  \frac{U_{\text{eff}}}{2}(n_{\tilde{f}} - 1)^2
\end{eqnarray}
eliminates the linear electron-phonon coupling, and renormalizes the onsite interaction $U_{\text{eff}} = U - 2\alpha \omega_0$ due to the phonon-induced reduction of the onsite interaction, assuming that the oscillator displaces by $x_0$. The Lang-Firsov transformation displaces the phonon operator such that the new phonon operator is,
\begin{eqnarray}\label{eq:displacedoscillator}
   e^{i\hat{p}x_0} b\dg e^{-i\hat{p}x_0} = \tilde b\dg  = b\dg + \sqrt{\alpha} \, (n_f - 1) .
\end{eqnarray}
The original f-creation operator now transforms to 
\begin{eqnarray}\label{eq:polaron}
   f\dg_{\sigma} \rightarrow e^{i\hat{p} x_0} f\dg_\sigma e^{-i \hat{p}x_0} =  D {\tilde f}\dg_{\sigma},
\end{eqnarray}
where $D = \exp{\left( \sqrt{\alpha} (b\dg-b )\right)} =\exp{\left( \sqrt{\alpha} (\tilde{b}\dg-\tilde{b} )\right)} $.  The operator ${\tilde f}\dg_{\sigma}$ creates the polaron, while  $D$ undresses the impurity from the polaron cloud to ensure $f\dg_\sigma$ creates the bare impurity electron.  Under this unitary transformation, the charge is unaltered, i.e $n_{\tilde f}=n_{f}$.  Henceforth, we will drop the tilde on the f operators, with the understanding that in the transformed Hamiltonian, the $f\dg $ are polaron creation operators.
and the impurity f-electron operator is dressed by the phonon cloud to form a polaronic operator.

Turning on the hybridization $V$, the changes in the phonon cloud structure for the polaron associated with fluctuations of the impurity valence can be absorbed in a redefinition of the hybridization. The full translated Anderson-Holstein Hamiltonian is,
\begin{eqnarray}\label{eq:transformedHammy}
    \tilde{H} &=& \sum_{\bk \sigma} \epsilon_{\bk} c\dg_{\bk \sigma} c_{\bk \sigma} + \omega_0(\tilde{b}\dg \tilde{b} +1/2) +  \frac{U_{\text{eff}}}{2}(n_f - 1)^2 + V \sum_{ \sigma} \left(c\dg_{ \sigma} f_{\sigma} D\dg +  f\dg_{\sigma} c_{\sigma} D\right). 
\end{eqnarray}
%
\section{Details of the Schrieffer Wolff Transformation for Strong Electron-Phonon Coupling and Derivation of $U^*$}\label{sec:details}
This section has two objectives. First, we demonstrate that the low-energy physics of the symmetric single impurity Anderson model coupled to slow phonons ($\omega_0 \ll U/2$)- where phonons are too slow to respond to charge fluctuations and thus do not affect correlated electron physics, as shown in section \ref{sec:slowphonons_analytic}- is governed by a conventional Kondo spin impurity Hamiltonian. In this regime, bare electrons form a local moment that gets screened by the conduction bath. Second, we establish the existence of a threshold electron-phonon coupling $\alpha^*$ above which the low-energy physics below scale $U^*$ transitions from a spin Kondo effect to a polaronic-induced frozen mixed valence phase.

We begin by considering the transformed Hamiltonian \eqref{eq:transformedHammy} at the extreme limit where the electron-phonon dimensionless coupling $\alpha_0$ where the effective onsite interaction vanishes $U_{\text{eff}} = U - 2\alpha_0 \omega_0 = 0$. It was shown in the main text (Sec. \ref{sec:slowphonons_analytic}) that the phonons are too slow to respond to form polarons, and hence the electronic degrees of freedom of the single impurity Anderson model are agnostic to the slow phonon. We now show using perturbation theory that this is indeed true within the Lang-Firsov transformation (see Appendix \ref{sec:langfirsovappendix}), even in the strong electron-phonon coupling limit where both $U$ and $\alpha_0$ is taken to infinity ($U \rightarrow \infty$, $\alpha_0 \rightarrow \infty$) in such a way to keep $U_{\text{eff}} = 0$. The Hamiltonian is then,
\begin{eqnarray}\label{eq:vanishingtransformedHammy}
    \tilde{H} &=& \sum_{\bk \sigma} \epsilon_{\bk} c\dg_{\bk \sigma} c_{\bk \sigma} + \omega_0(\tilde{b}\dg \tilde{b} +1/2) + V \sum_{ \sigma} \left(c\dg_{ \sigma} f_{\sigma} D\dg +  f\dg_{\sigma} c_{\sigma} D\right). 
\end{eqnarray}
In the atomic limit ($V = 0$) of this model, where the hybridization between the impurity and conduction bath is turned off, the ground state of the phonon bath is the unoccupied state $\vert \tilde{0}_b \rangle$ and the impurity has a quartet degeneracy.

Turning on the hybridization $V$ perturbatively involves displacing the atomic phonon ground state to the coherent states,
\begin{equation}
    \vert \tilde{+} \rangle = D \vert \tilde{0} \rangle, \;  \vert \tilde{-} \rangle = D\dg \vert \tilde{0} \rangle,
\end{equation}
when an electron tunnels from the impurity into the conduction bath, or vice versa. These phonon coherent states are high energy states with average energy $\alpha_0 \omega_0 = U/2$, where the equality stems from the condition that $U_{\text{eff}} = U - 2 \alpha_0 \omega_0 = 0$.

When the dimensionless electron-phonon coupling is large, the matrix elements of the coherent state displacement operators are exponentially suppressed
\begin{equation}
    \langle \tilde{0} \vert D \vert \tilde{0} \rangle = \langle \tilde{0}\vert D\dg \vert \tilde{0} \rangle = e^{-\frac{1}{2} \alpha_0} \approx 0 
\end{equation}
We will also ignore exponentially small matrix elements
\begin{equation}
    \langle \tilde{0} \vert D\vert \tilde{+} \rangle = \langle \tilde{0} \vert D\dg \vert \tilde{-} \rangle = e^{-2 \alpha_0} \approx 0.
\end{equation}
These two matrix elements constitute the mutual orthonormality condition between the coherent states $\{\vert \tilde{0}_b \rangle, \vert \tilde{-} \rangle, \vert \tilde{+} \rangle \}$.

To understand the low-energy physics of the impurity spin and charge sectors, we define the projectors into the low-energy spin sector,
\begin{equation}\label{eq:lowspinproj}
   P_{SL} = \left(\vert \uparrow_f \rangle \langle \uparrow_f \vert +  \vert \downarrow_f \rangle \langle \downarrow_f \vert \right) \otimes \vert \tilde{0} \rangle \langle \tilde{0} \vert,
\end{equation}
and the low energy charge sector,
\begin{equation}\label{eq:lowchargeproj}
   P_{CL} = \left(\vert 0_f \rangle \langle 0_f \vert +  \vert \uparrow \downarrow_f \rangle \langle \uparrow \downarrow_f \vert \right) \otimes \vert \tilde{0} \rangle \langle \tilde{0} \vert.
\end{equation}
We also define the projector to the high energy spin sector as,
\begin{eqnarray}\label{eq:highspinproj}
   P_{SH} &=& \left(\vert \uparrow_f \rangle \langle \uparrow_f \vert +  \vert \downarrow_f \rangle \langle \downarrow_f \vert \right) \otimes  \left(\vert \tilde{+} \rangle \langle \tilde{+} \vert + \vert \tilde{-} \rangle \langle \tilde{-} \vert\right),
\end{eqnarray}
and the projector to the high energy charge sector as,
\begin{eqnarray}\label{eq:highchargeproj}
   P_{CH} &=& \left(\vert 0_f \rangle \langle 0_f \vert +  \vert \uparrow \downarrow_f \rangle \langle \uparrow \downarrow_f \vert \right) \otimes  \left(\vert \tilde{+} \rangle \langle \tilde{+} \vert + \vert \tilde{-} \rangle \langle \tilde{-} \vert\right),
\end{eqnarray}
where the identity projection onto the conduction bath sector $\mathrm{1}_c$ has been suppressed for brevity.

Due to the mutual orthogonality of the coherent states $\{\vert\tilde{0}_b \rangle, \vert \tilde{-} \rangle, \vert \tilde{+} \rangle \}$, the low (high) energy spin and low (high) energy charge sectors decouple in the strong dimensionless electron-phonon coupling limit ($\alpha_0 \rightarrow \infty$). Next, we perform the Schrieffer-Wolff transformation to the low-energy impurity spin and charge sectors to show that the low-energy physics of the spin sector is governed by a spin Kondo model, and the low-energy physics of the charge sector is governed by a polaronic-induced frozen mixed valence phase.
\subsection{Low Energy Spin Sector}
\begin{equation}\label{eq:spinlowsectorappendix}
    H^{S} = \left[ 
\begin{array}{c|c} 
  H_{SL} & V\dg_{SL,CH} \\ 
  \hline 
  V_{CH,SL} &  H_{CH}
\end{array} 
\right], \quad V^S_{\text{mix}} = \left[ 
\begin{array}{c|c} 
0 & V\dg_{SL,CH} \\ 
  \hline 
  V_{CH,SL} &  0
\end{array} 
\right].
\end{equation}
\begin{eqnarray}
V\dg_{SL,CH} = P_{SL} \hat{V} P_{CH} = &V&\sum_{\beta} \left( c\dg_{\beta} \vert \bar{\beta}_f \rangle \langle \bar{\beta}_f \vert f_{\beta} \vert \uparrow \downarrow_f \rangle \langle \uparrow \downarrow_f \vert \otimes \vert \tilde{0} \rangle \langle \tilde{0} \vert D\dg \vert \tilde{+} \rangle \langle \tilde{+} \vert + \right. \cr  &&\left. \vert \beta_f \rangle \langle  \beta_f \vert f\dg_{\beta} \vert 0_f \rangle \langle 0_f \vert c_{\beta} \otimes \vert \tilde{0} \rangle \langle \tilde{0} \vert D \vert \tilde{-} \rangle \langle \tilde{-} \vert \right)
\end{eqnarray}
\begin{eqnarray}
   V_{CH,SL} = P_{CH} \hat{V} P_{SL} = &V&\sum_{\beta} \left( c\dg_{\beta} \vert 0_f \rangle \langle 0_f \vert f_{\beta} \vert \beta_f \rangle \langle \beta_f \vert \otimes \vert \tilde{-} \rangle \langle \tilde{-} \vert D\dg \vert \tilde{0} \rangle \langle \tilde{0} \vert + \right. \cr  &&\left. \vert \uparrow \downarrow_f \rangle \langle \uparrow \downarrow_f \vert f\dg_{\beta} \vert \bar{\beta}_f \rangle \langle \bar{\beta}_f \vert c_{\beta} \otimes \vert \tilde{+} \rangle \langle \tilde{+} \vert D \vert \tilde{0} \rangle \langle \tilde{0} \vert \right)
\end{eqnarray}
Performing the standard Schrieffer-Wolff to block diagonalize Eq. \ref{eq:spinlowsectorappendix}, 
\begin{eqnarray}
    H^S_{\text{eff}} = H_{SL} + \Delta H_{SL, SL'} = e^{S_s} H^S e^{-S_s} = H_{SL} + \frac{1}{2} P_{SL}[S_s, V^S_{\text{mix}}]P_{SL'}
\end{eqnarray}
we get the second order perturbative correction to the low energy Hamiltonian for the charge sector as,

\begin{eqnarray}\label{eq:changeHsL}
    \Delta H_{SL, SL'} &=& -\frac{V\dg_{SL,CH}V_{CH,lSL'}}{E^H_C - E^L_S} = -\frac{V\dg_{SL,CH}V_{CH,SL'}}{\alpha_0 \omega_0} = \frac{2V\dg_{SL,CH}V_{CH,SL'}}{U}\cr &=& -\frac{2V^2}{U} \sum_{\beta \gamma} \left(|\langle \tilde{+} \vert D \vert \tilde{0}_b \rangle |^2 c\dg_{\beta}\vert \bar{\beta}_f \rangle \langle \bar{\beta}_f \vert f_{\beta} \vert \uparrow\downarrow_f \rangle \langle \uparrow \downarrow_f \vert f\dg_{\gamma} \vert \bar{\gamma}_f \rangle \langle \bar{\gamma}_f \vert c_{\gamma} \otimes \vert \tilde{0} \rangle \langle \tilde{0} \vert + \right. \cr && \left. |\langle \tilde{-} \vert D\dg \vert \tilde{0}_b \rangle |^2 \vert \beta_f \rangle \langle \beta_f \vert f\dg_{\beta} \vert 0_f \rangle  c_\beta c\dg_{\gamma} \langle 0_f \vert f_{\gamma} \vert \gamma_f \rangle \langle \gamma_f \vert \otimes \vert \tilde{0} \rangle \langle \tilde{0} \vert \right) \cr &=& -\frac{2V^2}{U} \sum_{\beta \gamma} \left(c\dg_{\beta} f_{\beta} f\dg_{\gamma} c_{\gamma}  + f\dg_{\gamma} c_{\gamma} c\dg_{\beta} f_{\beta} \right)P_{n_f = 1}\otimes \vert \tilde{0} \rangle \langle \tilde{0} \vert. 
\end{eqnarray}
In the first line, we have used $\alpha_0 \omega_0 = U/2$. Using the Fierz identity $2 \delta_{\alpha \gamma} \delta_{\eta \beta} = \delta_{\alpha \beta} \delta_{\eta \gamma} + \vec{\sigma}_{\alpha \beta} \cdot \vec{\sigma_{\eta \gamma}}$, we can write the low energy effective Hamiltonian for the local moment coupled to the bath as a Kondo impurity model,
\begin{equation}
    H^S_{\text{eff}} =  \sum_{\bk \sigma} \epsilon_{\bk} c\dg_{\bk \sigma} c_{\bk \sigma} + \Delta H_{SL, SL'} = \sum_{\bk \sigma} \epsilon_{\bk} c\dg_{\bk \sigma} c_{\bk \sigma} +\underbrace{\frac{4V^2}{U}}_{J_K} \vec{\sigma}(0)\cdot \vec{S}_f,
\end{equation}
where we have dropped the residual potential scattering term which vanishes for the particle-hole symmetric case, $J_K$ is the Kondo coupling. This model describes the low energy physics of the bare repulsive single impurity Anderson.
\\
\subsection{Low Energy Charge Sector}
\begin{equation}\label{eq:chargelowsectorappendix}
    H^{C} = \left[ 
\begin{array}{c|c} 
  H_{CL} & V\dg_{CL,SH} \\ 
  \hline 
  V_{SH,CL} &  H_{SH}
\end{array} 
\right], \quad V^C_{\text{mix}} = \left[ 
\begin{array}{c|c} 
0 & V\dg_{CL,SH} \\ 
  \hline 
  V_{SH,CL} &  0
\end{array} 
\right].
\end{equation}
\begin{eqnarray}
   V\dg_{CL,SH} = P_{CL} \hat{V} P_{SH} = &V&\sum_{\beta} \left( c\dg_{\beta} \vert 0_f \rangle \langle 0_f \vert f_{\beta} \vert \beta_f \rangle \langle \beta_f \vert \otimes \vert \tilde{0} \rangle \langle \tilde{0} \vert D\dg \vert \tilde{+} \rangle \langle \tilde{+} \vert + \right. \cr  &&\left. \vert \uparrow \downarrow_f \rangle \langle  \uparrow \downarrow_f \vert f\dg_{\beta} \vert \bar{\beta}_f \rangle \langle \bar{\beta}_f \vert c_{\beta} \otimes \vert \tilde{0}_\rangle \langle \tilde{0} \vert D \vert \tilde{-} \rangle \langle \tilde{-} \vert \right)
\end{eqnarray}
\begin{eqnarray}
   V_{SH,CL} = P_{SH} \hat{V} P_{CL} = &V&\sum_{\beta} \left( c\dg_{\beta} \vert \bar{\beta}_f \rangle \langle \bar{\beta}_f \vert f_{\beta} \vert \uparrow \downarrow_f \rangle \langle \uparrow \downarrow_f \vert \otimes \vert \tilde{-} \rangle \langle \tilde{-} \vert D\dg \vert \tilde{0} \rangle \langle \tilde{0} \vert + \right. \cr  &&\left. \vert \beta_f \rangle \langle  \beta_f \vert f\dg_{\beta} \vert 0_f \rangle \langle 0_f \vert c_{\beta} \otimes \vert \tilde{+} \rangle \langle \tilde{+} \vert D \vert \tilde{0} \rangle \langle \tilde{0} \vert \right)
\end{eqnarray}
Performing the standard Schrieffer-Wolff to block diagonalize Eq. \ref{eq:chargelowsectorappendix}, 
\begin{eqnarray}
    H^C_{\text{eff}} = H_{CL} + \Delta H_{CL, CL'} = e^{S_c} H^C e^{-S_c} = H_{CL} + \frac{1}{2} P_{CL} [S_c, V^C_{\text{mix}}]P_{CL'}
\end{eqnarray}
we get the second order perturbative correction to the low energy Hamiltonian for the charge sector as,

\begin{eqnarray}\label{eq:changeHcL}
    \Delta H_{CL, CL'} &=& -\frac{V\dg_{CL,SH}V_{SH,CL'}}{E^H_S - E^L_C} = -\frac{V\dg_{CL,SH}V_{SH,CL'}}{\alpha_0 \omega_0}  = -\frac{2V\dg_{CL,SH}V_{SH,CL'}}{U} \cr &=& -\frac{2V^2}{U} \sum_{\beta \gamma} \left(|\langle \tilde{+} \vert D \vert \tilde{0} \rangle |^2 c\dg_{\beta}\vert 0_f \rangle \langle 0_f \vert f_{\beta} \vert \beta_f \rangle \delta_{\beta \gamma} \langle \gamma_f \vert f\dg_{\gamma} \vert 0_f \rangle \langle 0_f \vert c_{\gamma} \otimes \vert \tilde{0} \rangle \langle \tilde{0} \vert + \right. \cr && \left. |\langle \tilde{-} \vert D\dg \vert \tilde{0} \rangle |^2 \vert \uparrow \downarrow_f \rangle \langle \uparrow \downarrow_f \vert f\dg_{\beta} \vert \bar{\beta}_f \rangle \delta_{\beta \gamma} c_\beta c\dg_{\gamma} \langle \bar{\gamma}_f \vert f_{\gamma} \vert \uparrow \downarrow_f \rangle \langle \uparrow \downarrow_f \vert \otimes \vert \tilde{0} \rangle \langle \tilde{0} \vert \right) \cr &=& -\frac{2V^2}{U} \sum_{\beta} \left(c\dg_{\beta} f_{\beta} f\dg_{\beta} c_{\beta} P_{n_f = 0} + f\dg_{\beta} c_{\beta} c\dg_{\beta} f_{\beta} P_{n_f = 2} \right)\otimes \vert \tilde{0} \rangle \langle \tilde{0} \vert .
\end{eqnarray}
Hence the low energy effective Hamiltonian for the impurity charge sector is,
\begin{equation}
    H^C_{\text{eff}} = \sum_{\bk \sigma} \epsilon_{\bk} c\dg_{\bk \sigma} c_{\bk \sigma} + \Delta H_{CL, CL'} = \sum_{\bk \sigma} \epsilon_{\bk} c\dg_{\bk \sigma} c_{\bk \sigma}  -\frac{2V^2}{U} \sum_{\beta} \left(c\dg_{\beta} f_{\beta} f\dg_{\beta} c_{\beta} P_{n_f = 0} + f\dg_{\beta} c_{\beta} c\dg_{\beta} f_{\beta} P_{n_f = 2} \right)\otimes \vert \tilde{0} \rangle \langle \tilde{0} \vert, 
\end{equation}
which has ``Ising'' fluctuations.
\end{widetext}
\end{document}